\documentclass[11pt]{article}

\usepackage[usenames,dvipsnames]{pstricks}
\usepackage{epsfig}
\usepackage{pst-grad} 
\usepackage[space]{grffile} 
\usepackage{etoolbox} 
\makeatletter 
\patchcmd\Gread@eps{\@inputcheck#1 }{\@inputcheck"#1"\relax}{}{}
\makeatother

\usepackage{graphicx}
\usepackage{amsmath}
\usepackage{psfrag}
\usepackage{color,soul}
\usepackage{natbib}
\usepackage{float}
\usepackage{overpic}
\usepackage{tikz}
\usepackage{multirow}
\usepackage{bm}
\usepackage{textgreek}
\usepackage{url}
\usepackage{amssymb}
\usepackage{xfrac}
\usepackage{setspace}
\usepackage{gensymb}
\usepackage{mathtools}
\usepackage{arydshln}
\usepackage{MnSymbol} 

\def\NoNumber#1{{\def\alglinenumber##1{}\State #1}\addtocounter{ALG@line}{-1}}
\usepackage{booktabs} 
\usepackage{algorithm}
\usepackage[noend]{algpseudocode}

\def \eg{\emph{e.g.}, }
\def \ie{\emph{i.e.}, }
\def \Dt{$\Delta t$~}

\usepackage{hyperref}
\definecolor{darkblue}{rgb}{0,0,0.80}
\hypersetup{
    colorlinks=true,
    linkcolor={darkblue},
    citecolor={darkblue},
    filecolor=black,
    urlcolor={darkblue},
    plainpages=false,
}

\usepackage[noblocks]{authblk}
\usepackage[margin=1in]{geometry}
\usepackage[title]{appendix}

\newcommand\affiliation[1]{\gdef\@affiliation{\let\aff\aff@inst#1}}
\gdef\@affiliation{}
\def\email#1{Email address for correspondence: #1}
\def\aff#1{\ignorespaces\textsuperscript{#1}}
\def\corresp#1{\unskip\thanks{#1}}
\numberwithin{equation}{section}

\renewenvironment{abstract}
{\begin{quote}
\noindent \rule{\linewidth}{.5pt}\par{\bfseries \abstractname.}}
{\medskip\noindent \rule{\linewidth}{.5pt}
\end{quote}
}

  \DeclareTextFontCommand\textsfi{\usefont{OT1}{cmss}{m}{sl}}
  \DeclareMathAlphabet\mathsfi            {OT1}{cmss}{m}{sl}
  \DeclareTextFontCommand\textsfb{\usefont{OT1}{cmss}{bx}{n}}
  \DeclareMathAlphabet\mathsfb            {OT1}{cmss}{bx}{n}
  \DeclareTextFontCommand\textsfbi{\usefont{OT1}{cmss}{m}{sl}}
  \DeclareMathAlphabet\mathsfbi            {OT1}{cmss}{m}{sl}
  \IfFileExists{t1phv.fd}
  {\DeclareTextFontCommand\textsfbi{\usefont{T1}{phv}{b}{it}}
  \DeclareMathAlphabet\mathsfbi            {T1}{phv}{b}{it}
  }{}
  \IfFileExists{ot1phv.fd}
  {\DeclareTextFontCommand\textsfbi{\usefont{OT1}{phv}{b}{it}}
  \DeclareMathAlphabet\mathsfbi            {OT1}{phv}{b}{it}
  }{}
  
\DeclareSymbolFont{matha}{OML}{txmi}{m}{it}

\newcommand{\edit}[1]{\color{black}#1 \color{black}}

\title{\bf Efficient harmonic resolvent analysis via time-stepping}

\author[1]{\bf Ali Farghadan}
\author[1]{\bf Jung Jung}
\author[1]{\bf Rutvij Bhagwat}
\author[1]{\bf Aaron Towne\corresp{\email{towne@umich.edu}}}

\affil[1]{\normalsize Department of Mechanical Engineering, University of Michigan, Ann Arbor, MI, USA }

\date{}
\begin{document}
\maketitle

\begin{abstract}

We present an extension of the RSVD-\Dt algorithm initially developed for resolvent analysis of statistically stationary flows to handle harmonic resolvent analysis of time-periodic flows. The harmonic resolvent operator, as proposed by \citet{Padovanetal20}, characterizes the linearized dynamics of time-periodic flows in the frequency domain, and its singular value decomposition reveals forcing and response modes with optimal energetic gain. However, computing harmonic resolvent modes poses challenges due to $(i)$ the coupling of all $N_{\omega}$ retained frequencies into a single harmonic resolvent operator and $(ii)$ the singularity or near-singularity of the operator, making harmonic resolvent analysis considerably more computationally expensive than a standard resolvent analysis. To overcome these challenges, the RSVD-\Dt algorithm leverages time stepping of the underlying time-periodic linearized Navier-Stokes operator, which is $N_{\omega}$ times smaller than the harmonic resolvent operator, to compute the action of the harmonic resolvent operator. We develop strategies to minimize the algorithm's CPU and memory consumption, and our results demonstrate that these costs scale linearly with the problem dimension. We validate the RSVD-\Dt algorithm by computing modes for a periodically varying Ginzburg-Landau equation and demonstrate its performance using the flow over an airfoil.

\end{abstract}


\section{Introduction}

Model reduction plays a critical role in the study of fluid mechanics due to the complex and high-dimensional nature of fluid flows, especially turbulent ones. In particular, modal decompositions, both data-driven and equation-based, have proven to be effective at identifying low-dimensional sets of modes associated with interpretable coherent structures that significantly influence quantities of engineering interest such as kinetic energy, heat and mass transfer, and noise emissions \citep{Tairaetal17, Towneetal18}. In particular, resolvent analysis has been extensively employed to comprehend, model, and control statistically stationary turbulent flows such as wall-bounded turbulence \citep{DawsonMcKeon19}, turbulent boundary layer \citep{SippMarquet13}, and jet flows \citep{Pickeringetal21noise}. However, real-world fluid systems often exhibit non-stationary behavior, including periodicity, rendering resolvent analysis less effective. Examples of flows exhibiting periodic behavior include internal waves in a stratified fluid \citep{TroyKoseff05}, vortex shedding in the wake of a bluff body \citep{Giannenasetal22}, and pulsatile blood flow \citep{FarghadanArzani19}. Recently, an extension of resolvent analysis, known as harmonic resolvent analysis, has been developed by \citet{Padovanetal20}. This extension characterizes the perturbation forcing and responses around a periodic base flow, providing a more effective tool for analyzing, modeling, and controlling periodically time-varying flows. The harmonic resolvent operator can be considered a special case of the harmonic transfer function introduced by \citet{Wereley90}, wherein the operator exhibits a dominant frequency.

Investigating the harmonic resolvent operator is one way to identify the optimal spatio-temporal perturbations and their corresponding responses in fluid systems. Unlike the resolvent operator, the harmonic resolvent operator takes into consideration the periodicity of the base flow by representing it in the form of a Fourier series. This allows for characterization of the dynamic behavior of the system, particularly with respect to its spectral components and it provides a means to understand and explore the interactions between input and output modes through the linearized dynamics. Harmonic resolvent analysis determines the triadic interactions between the input frequency $\omega_1$, the base flow frequency $\omega_1 - \omega_2$, and the response frequency $\omega_2$. The number of Fourier modes of the base flow is determined by its power spectral density (PSD), which in turn determines the number of triads interacting with the forcing. Typically, periodic flows with a dominant periodicity exhibit a multimodal nature with their most energetic modes being harmonics of the base frequency. When the base flow is statistically stationary, linearization around the temporal average is sufficient, and traditional resolvent analysis arises as a special case.

Recent studies have expanded on this research area to model and control the dynamics of turbulent flows with periodic motions. \citet{Linetal23} applied the harmonic resolvent framework for linear time-periodic systems with more than one dominant frequency, offering a tool to study the flow control of a plunging cylinder. In a distinct approach, \citet{Franceschinietal22} proposed a novel method aimed at capturing the phase dependency of small-scale turbulent phenomena in flows exhibiting a periodic limit cycle. This method builds upon classical spectral proper orthogonal decomposition (SPOD) and resolvent analyses but introduces a quasi-steady approximation that effectively separates the high-frequency turbulent component from the slow periodic oscillation. Moreover, \citet{HeidtColonius23} have pioneered the theory and algorithm for cyclostationary SPOD (CS-SPOD). This innovative approach extends the conventional SPOD algorithm to effectively capture the characteristics of flows characterized by periodic motions. CS-SPOD is closely tied to harmonic resolvent analysis, akin to the relationship between SPOD and resolvent analysis \citep{Towneetal18}.


\edit{Harmonic resolvent analysis offers a promising avenue for new discoveries in physics.  For example, recent research conducted by \citet{Wuetal22} explored the response of a turbulent separation bubble to zero-net-mass-flux actuation, with a base flow characterized by periodicity. The noteworthy finding revealed an agreement between the reduction in separation bubble size identified through both harmonic resolvent optimal response and direct numerical simulation. Another contribution in this realm comes from the work of \citet{PadovanRowley22}, who investigated the driving mechanisms underlying vortex pairing in an incompressible forced axisymmetric jet. They showed that the optimal forcing of the subharmonic resolvent operator featured spatiotemporal structures leading to the nonlinear vortex pairing phenomenon. In general, harmonic resolvent analysis can facilitate the understanding of energy transfer mechanisms \citep{Jinetal21} and is a potentially powerful tool for secondary stability analyses involving periodic additions to the base flow, with applications such as boundary layer stability and transition \citep{Herbert84}, screeching jets \citep{Edgingtonetal21}, and the effect of mass injection on secondary instability \citep{KumarPrakash22}.}

From a computational standpoint, computing harmonic resolvent modes presents similar challenges to resolvent analysis, including the need to invert sparse matrices in Fourier space. This challenge is further exacerbated in harmonic resolvent analysis due to expanded system size, where all relevant frequencies are merged into a unified operator, in contrast to resolvent analysis, where each frequency has its own independent system. The second shared challenge involves performing the singular value decomposition (SVD). \edit{Turbulent and other complex flows characterized by periodic motions require high-resolution meshes and potentially large computational domains. This results in a sizable linearized operator, introducing formidable computational challenges when it comes to matrix inversion and performing SVD.} A modified randomized SVD (RSVD) can be used to simultaneously overcome both computational challenges \citep{Moarrefetal13, Ribeiroetal20, Padovanetal20}. Nevertheless, the modified RSVD, referred to as ``RSVD-LU'' in this study, becomes impractical for tackling large-scale turbulent flows since it requires the lower-upper (LU) decomposition of the harmonic resolvent operator. Recently, RSVD-$\Delta t$, an improvement to RSVD-LU, addresses bottlenecks in resolvent analysis \citep{Farghadanetal23}. The approach of time-stepping, initially developed by \citet{Monokrousosetal10} and further optimized by \citet{Martinietal21}, allows for computing the action of an operator on a vector without solving a linear system in Fourier space, making it a key component of RSVD-$\Delta t$. 

The main objective of this study is to expand the applicability of the RSVD-\Dt algorithm, initially crafted for resolvent analysis, to facilitate the computation of harmonic resolvent modes. A streaming approach is adopted for time integration, reducing memory usage while maintaining computational efficiency. Moreover, an efficient transient removal strategy is introduced, cutting the temporal length of the integration and thereby minimizing the CPU cost of RSVD-$\Delta t$ for achieving a desired accuracy. The RSVD-\Dt algorithm for periodic flows maintains linear scalability in both CPU time and memory consumption. This scalability enables the efficient computation of harmonic resolvent modes for large-scale turbulent flows that were previously out of reach.

The structure of our paper unfolds as follows. We begin by explaining resolvent analysis in $\S$\ref{sec:res}, followed by the formulation of harmonic resolvent analysis in $\S$\ref{sec:hres}, where the computation of harmonic resolvent modes using RSVD-LU is briefly described. The RSVD-\Dt algorithm is introduced in $\S$\ref{sec:time-stepping}. A comparison of the computational complexity of algorithms and suggested improvements appear in $\S$\ref{sec:computation}. Further details on performance and error analysis are presented in $\S$\ref{sec:Opt}. In $\S$\ref{sec:TestCases}, we conduct two test cases to demonstrate the accuracy and cost efficiency of RSVD-\Dt for two periodic systems, and we conclude with final remarks in $\S$\ref{sec:conclusion}.

\section{Resolvent analysis} \label{sec:res}

Before considering the time-periodic case, we briefly review standard resolvent analysis for stationary flows. The Navier-Stokes equations can be linearized around a base flow via Reynolds decomposition of the state, $\bm q = \bar{\bm q} + \bm q'$, where $\bar{\bm q}$ represents the temporal average of the flow, and $\bm q'$ denotes the time-varying fluctuations. The dynamics of perturbations are then described by the linear time-invariant system
\begin{equation}
\frac{d\bm q'}{dt} = \bm A(\bar{\bm q})\bm q' + \bm f'(\bar{\bm q} , \bm q'),
\label{eqn:linsys}
\end{equation}
where $\bm A \in \mathbb{C}^{N \times N}$ is the linearized Navier-Stokes (LNS) operator, $N$ is the state dimension of the discretized system, and $\bm f'$ represents an external forcing as well as nonlinear terms that are treated as an additional forcing, as suggested by \citet{MckeonSharma10}. It should be noted that $\bm A$ is time-independent because the base flow is time-invariant. By taking the Fourier transform
\begin{equation}
\mathcal{F}(\cdot) = \hat{(\cdot)}(\omega) = \int_{-\infty}^{+\infty} (\cdot)e^{-i\omega t} \,dt
\label{eqn:FT}
\end{equation}
in time, we obtain
\begin{equation}
\hat{\bm q}(\omega) = \bm R(\omega)\hat{\bm f}(\omega),
\label{eqn:resolvent_eqn}
\end{equation}
where $\omega$ represents the frequency and $\hat{(\cdot)}$ denotes the frequency-domain representation of the time-domain vector. The resolvent operator
\begin{equation}
\bm R = (\text{i}\omega \bm I - \bm A)^{-1} \in \mathbb{C}^{N \times N}
\label{eqn:resolvent_op}
\end{equation} 
is a transfer function that maps the input forcing to the output response. Here, $\text{i} = \sqrt{-1}$ and $\bm I$ is the identity matrix of appropriate dimension.

To analyze the perturbation characteristics of a system, the transfer function can be examined across a range of frequencies. Typically, the most amplified responses are identified by computing the left singular vectors $\bm U_R$ of the resolvent operator,
\begin{equation}
\bm R = \bm U_R \boldsymbol{\varSigma}_R \bm V_R^*,
\label{eqn:res_svd}
\end{equation} 
where $(\cdot)^*$ denotes complex conjugate transpose. The optimal forcing inputs that lead to the highest amplifications comprise the right singular vectors $\bm V_R$, and the degree of amplifications are determined by the singular values $\boldsymbol{\varSigma}_R$. 

The SVD in \ref{eqn:res_svd} implies that disturbance amplitudes are measured in a Euclidean norm. More generally, $||\boldsymbol{x}||^2_f = \langle \boldsymbol{x}, \boldsymbol{x} \rangle_f = \boldsymbol{x}^*\bm W_f\boldsymbol{x}$ computes the $f$-norm of any vector $\boldsymbol{x}$, and input and output norms can be different, \ie $||\cdot||_q = ||\cdot||_f$ is not necessary. Additionally, one can define an input matrix $\bm B$ to restrict the forcing and an output matrix $\bm C$ to extract the output of interest from the state. For the sake of notational brevity, we assumed the weight, input, and output matrices to be the identity, \ie $\bm W = \bm B = \bm C = \bm I$, throughout this paper. However, our algorithm, with minor adjustments, can accommodate non-identity weight, input, and output matrices by computing the modes of the modified resolvent operator 
\begin{equation}
\tilde{\bm R} = \bm W^{1/2}_q \bm C \bm R \bm B \bm W^{-1/2}_f.
\label{eqn:mod_res_op}
\end{equation}
For more details, please refer to \citet{Farghadanetal23}.

\section{Harmonic resolvent analysis} \label{sec:hres}

In this section, we provide an overview of harmonic resolvent analysis. After reviewing its basic formulation, we discuss the singularity issue that can arise in some special flows, the truncation of the harmonic resolvent operator, and two variations of harmonic resolvent analysis, namely cross-frequency amplification and subharmonic resolvent. Lastly, a brief overview is provided for the state-of-the-art algorithm used to compute the harmonic resolvent modes and gains.

\subsection{Formulation} \label{sec:formulation}

Analogous to resolvent analysis, by inserting the generalized Reynolds decomposition $\bm q(t) = \bar{\bm q}(t) + \bm q^{\prime}(t)$ into the Navier-Stokes equations, the linearization around a periodic base flow $\bar{\bm q}(t) = \bar{\bm q}(t+T)$ leads to the linear time-periodic system
\begin{equation}
\frac{d\bm q'}{dt} = \bm A_p\bm q' + \bm f,
\label{eqn:linsys_harmonic}
\end{equation}
where $\bm A_p(t) = \bm A_p(t + T) \in \mathbb C^{N\times N}$ is the periodic LNS operator, $\omega_f$ is the fundamental frequency, and $T = 2\pi/\omega_f$ is defined as the fundamental cycle length. As both $\bm A_p$ and $\bm q'$ are time-periodic, the Fourier transform of \eqref{eqn:linsys_harmonic} incorporates interactions between all pairs of frequencies. For simplicity, we drop the prime notation for fluctuations from here on out.

Expanding $\bm A_p(t)$ and $\bm q(t)$ in terms of the Fourier series
\begin{equation}
\begin{gathered}
\bm A_p(t) = \sum_{j = -\infty}^{\infty} \hat{\bm A}_{p, j} e^{\text{i} j\omega_f t}, 
\\
\bm q(t) = \sum_{j = -\infty}^{\infty} \hat{\bm q}_j e^{\text{i} j\omega_f t}, 
\label{eqn:periodic_A_q}
\end{gathered}
\end{equation}
where $(\cdot)_j$ denotes $j^{th}$ harmonic of $\omega_f$, and substituting these expansions into \eqref{eqn:linsys_harmonic}, we obtain 
\begin{equation}
\left[\bm T\hat{\bm q}\right]_k = \text{i}k\omega_f \hat{\bm q}_k - \sum_{j = -\infty}^{\infty} \hat{\bm A}_{k - j} \hat{\bm q}_j = \hat{\bm f}_k.
\label{eqn:inverse_harmonic_res}
\end{equation}
Here, $\bm T$ is an infinite dimensional block matrix of the form
\begin{equation}
\bm T = 
\begin{bmatrix} 
    \ddots & \vdots  & \vdots & \vdots & \vdots & \udots \\
    \dots & \edit{\bm L_{-1}} & -\hat{\bm A}_{-1}  & -\hat{\bm A}_{-2} & -\hat{\bm A}_{-3} & \dots \\
    \dots & -\hat{\bm A}_1  &  \edit{\bm L_{0}}  & -\hat{\bm A}_{-1} & -\hat{\bm A}_{-2} & \dots \\
    \dots & -\hat{\bm A}_2  & -\hat{\bm A}_1 &  \edit{\bm L_{1}} & -\hat{\bm A}_{-1} & \dots \\
    \dots & -\hat{\bm A}_3  & -\hat{\bm A}_2 & -\hat{\bm A}_1 &  \edit{\bm L_{2}}  & \dots \\
    \udots & \vdots  & \vdots & \vdots & \vdots & \ddots 
\end{bmatrix},
\end{equation}
where the diagonal entries are in fact the inverse of resolvent operators at various frequencies,
\begin{equation}
\edit{\bm L_j =} \bm R_{j}^{-1} = \text{i} j\omega_f \bm I - \hat{\bm A}_{p, 0}, \forall j \in \mathbb{Z}.
\label{eqn:resolvent_enteries}
\end{equation}
The harmonic resolvent operator 
\begin{equation}
\bm H = \bm T^{-1}
\label{eqn:Hres}
\end{equation}
transfers the harmonic inputs to the outputs for periodic base flows, 
\begin{equation}
\hat{\bm q} = \bm H\hat{\bm f}.
\label{eqn:Hres_eqn}
\end{equation}

The SVD of the harmonic resolvent operator
\begin{equation}
\bm H = \textbf{U}_H \boldsymbol{\Sigma}_H \textbf{V}_H^*
\label{eqn:Hres_svd}
\end{equation} 
unveils the most amplified responses $\textbf{U}_H$ that correspond to optimal forcing $\textbf{V}_H$, each accompanied by associated amplification magnitudes $\boldsymbol{\Sigma}_H$. The matrix $\bm H$ encompasses the base flow, the fundamental frequency, and higher harmonics. 
Whereas individual frequencies can be analyzed separately in a standard resolvent analysis, all frequencies are coupled within $\bm H$ in harmonic resolvent analysis and must be considered simultaneously. Furthermore, the modes of the modified harmonic resolvent operator, defined as
\begin{equation}
\tilde{\bm H} = \bm W^{1/2}_q \bm C \bm H \bm B \bm W^{-1/2}_f,
\label{eqn:mod_Hres_op}
\end{equation}
can be computed using our algorithm with the same minor adjustments as detailed in \citet{Farghadanetal23}. In a special case where the base flow is time-independent, \ie $\hat{\bm A}_{p, j} = 0$ for $j \ne 0$, $\bm H$ reduces to
\begin{equation}
\bm H = 
\begin{bmatrix} 
    \ddots & \vdots  & \vdots & \vdots & \udots \\
    \dots & \bm R_{-1} & 0  & 0 & \dots \\
    \dots & 0  &  \bm R_0  & 0 & \dots \\
    \dots & 0  & 0 &  \bm R_1 & \dots \\
    \udots & \vdots  & \vdots & \vdots & \ddots 
\end{bmatrix},
\label{eqn:Hres_esp}
\end{equation}
so each frequency is decoupled. Therefore, resolvent analysis can be perceived as a special case of harmonic resolvent analysis.


\subsection{Singular or near-singular nature of \texorpdfstring{$\bm T$}{T}} \label{sec:singularity}

The operator $\bm T$ tends to be ill-conditioned for many periodic systems \citep{Padovanetal20}. A system is considered ill-conditioned when the ratio of the largest and smallest singular values is large $(\kappa = \sigma_{\text{max}}/\sigma_{\text{min}} \sim O(10^4)$ or higher). The smallest singular value of $T$ can be shown to reach machine precision zero if the periodic base flow satisfies the Navier-Stokes equations \citep{Padovanetal20, LeclercqSipp23}, \ie when
\begin{equation}
\frac{d\bm{\bar q}}{dt} = \boldsymbol{\mathcal N}(\bm{\bar q}),
\label{eqn:NS}
\end{equation}
where $\boldsymbol{\mathcal N}$ is the nonlinear Navier-Stokes operator. Taking a time derivative of both sides,
\begin{equation}
\frac{d\left(\frac{d\bm{\bar q}}{dt}\right)}{dt} = \frac{d\left(\boldsymbol{\mathcal N}(\bm{\bar q})\right)}{dt} = \frac{\partial \boldsymbol{\mathcal N}}{\partial \bm{\bar q}} \frac{d\bm{\bar q}}{dt} = \bm A_p\frac{d\bm{\bar q}}{dt}.
\label{eqn:NS_timeDerivative}
\end{equation}
Taking a Fourier transform of \eqref{eqn:NS_timeDerivative} and recalling that $\bm T$ is obtained as shown in \eqref{eqn:inverse_harmonic_res}, $\widehat{d\bm{\bar q}/dt}$ is a non-trivial solution of 
\begin{equation}
\bm T \bm q = 0,
\label{eqn:Null}
\end{equation}
residing within the null space of $\bm T$, thereby proving the singularity of $\bm T$. In systems where the base flow approximately satisfies \eqref{eqn:NS}, $\bm T$ becomes nearly singular. Consequently, irrespective of the specific periodic system under consideration, $\bm T$ might be poorly conditioned, presenting even more challenges when dealing with large systems.

\citet{Padovanetal20} demonstrated that defining the harmonic resolvent operator in a manner that eliminates the phase shift imparted by the Fourier coefficients of $d\bm{\bar q}/dt$ is effective, without adversely affecting the other dominant amplification mechanisms. By constraining $\bm T$ to a subspace $\bm U^{\perp}$ that is orthogonal to the direction of the phase shift $\widehat{d\bm{\bar q}/dt}$, the range of the harmonic resolvent operator is limited to $\bm U^{\perp}$. The key concept here is to eliminate the singular vectors associated with the smallest singular value of $\bm T$ without affecting the other singular values and vectors of $\bm T$. Given that the phase shift resides in the null space of $\bm T$, the desired right singular vector $\bm v = \widehat{\frac{d\bm{\bar q}}{dt}}/||\widehat{\frac{d\bm{\bar q}}{dt}}||$ is readily available. We can also solve 
\begin{equation}
\bm T^* \bm{\tilde u} = \bm v
\label{eqn:compute_u}
\end{equation}
and compute the corresponding left singular vector as $\bm u = \bm{\tilde u}/||\bm{\tilde u}||$. Upon obtaining $\bm v$ and $\bm u$, the process becomes straightforward: projecting out these modes from the forcing and response terms effectively restricts $\bm T$ to $\bm U^{\perp}$. These steps are elucidated in detail in \citet{Padovanetal20}. 

To compute the action of the harmonic resolvent operator on a given vector (or matrix), our goal is to determine the vector (or matrix) $\bm q$ that satisfies $\bm q = \bm H \bm f$. In brief, to compute the action of a singular $\bm H$ on a vector $\tilde{\bm f}$, the component along $\bm u$ must be projected out as 
\begin{equation}
\tilde{\bm f}_{in} = \tilde{\bm f} - \bm u(\bm u^*\tilde{\bm f}),
\label{eqn:f_in}
\end{equation}
and this will provide a response 
\begin{equation}
\bm T \bm{\tilde q} =  \tilde{\bm f}_{in} \rightarrow \bm{\tilde q} =  \bm H \tilde{\bm f}_{in},
\label{eqn:q_in}
\end{equation}
which is orthogonal to the direction of the phase shift. Nevertheless, to mitigate potential round-off errors, it is advisable to refine the output by projecting out the component along $\bm v$ as
\begin{equation}
\tilde{\bm q}_{out} = \bm{\tilde q} - \bm v(\bm v^*\bm{\tilde q}).
\label{eqn:q_out}
\end{equation}
Similarly, when computing the action of singular $\bm H^*$ on a vector, the initial step involves projecting out the component along $\bm v$. After the response is obtained, to mitigate round-off errors, the component along $\bm u$ is projected out.

\subsection{Truncating the harmonic resolvent operator} \label{sec:truncation}

Computing harmonic resolvent modes is contingent upon $\bm T$ being of finite size. The size and block sparsity pattern of $\bm T$ are determined by two key variables: the number of frequencies within the base flow, $N_b$, and within the response, $N_{\omega}$. Here, we elucidate the impact of each variable on the operator.

The rows of $\bm T$ consist of $N_b$-stencil blocks, dictated by the presence of non-zero elements in $\hat{\bm A}_{p, j}$. In the context of periodic flows, the time-dependent base flow can be expressed using a Fourier series expansion
\begin{equation}
\bar{\bm q}(t) = \sum_{j\omega_f \in \Omega_{\bar q}} \hat{\bar{\bm q}}_je^{\text{i}j\omega_f t}.
\label{eqn:periodic_mean}
\end{equation}
The set $\Omega_{\bar q}$ encompasses the relevant frequencies associated with the dominant flow structures. Typically, $\Omega_{\bar q}$ is a subset of $j \omega_f$ with $j \in \mathbb{Z}$ and encapsulates the majority of energy within the periodic flows. By retaining the base periodicity and a limited number of higher harmonics, it becomes possible to simplify the base flow representation, resulting in $N_b$ frequencies within $\Omega_{\bar q}$ and $N_b$ non-zero elements within each row of $\bm T$.

On the other hand, we aim to resolve perturbations occurring at temporal frequencies $\omega \in \Omega_{q}$, where $\Omega_{q}$ represents a subset of $j \omega_f$ with $j \in \mathbb{Z}$, although the specific choice of harmonics for $\Omega_{q}$ may vary, as further elucidated in $\S$\ref{sec:sub}. Typically, $\Omega_{\bar q}$ is a subset of $\Omega_{q}$, \ie the perturbation frequency content spans a wider range of harmonics \citep{Padovanetal20}. The number of frequencies $N_{\omega}$ expanding both $\bm f$ and $\bm q$ determines the dimensions of the block matrices, yielding a size of $(N_{\omega}N) \times (N_{\omega}N)$ for $\bm T$.

We provide a simple example to visually observe the harmonic resolvent operator. Suppose we use $\Omega_{\bar q} = \{-2, -1, 0, 1, 2\}\omega_f$ to expand the LNS operator and $\Omega_{q} = \{-4, -3, \dots, 3, 4\}\omega_f$ to expand the perturbations. In this case, the stencil length would be $N_b = 5$, and the size of the block matrix would be $(9N)\times(9N)$. The harmonic resolvent operator is then
\begin{equation}
\bm H = 
\begin{bmatrix} 
    \edit{\bm L_{-4}} & -\hat{\bm A}_{-1} & -\hat{\bm A}_{-2}  & 0 & 0 & 0 & 0  & 0 & 0 \\
    -\hat{\bm A}_{1}  & \edit{\bm L_{-3}} & -\hat{\bm A}_{-1}  & -\hat{\bm A}_{-2} & 0 & 0 & 0  & 0 & 0 \\
    -\hat{\bm A}_{2} & -\hat{\bm A}_{1}  & \edit{\bm L_{-2}} & -\hat{\bm A}_{-1}  & -\hat{\bm A}_{-2} & 0 & 0 & 0  & 0 \\
    0 & -\hat{\bm A}_{2} & -\hat{\bm A}_{1}  & \edit{\bm L_{-1}} & -\hat{\bm A}_{-1}  & -\hat{\bm A}_{-2} & 0 & 0 & 0 \\
    0 & 0 & -\hat{\bm A}_{2} & -\hat{\bm A}_{1}  & \edit{\bm L_{0}} & -\hat{\bm A}_{-1}  & -\hat{\bm A}_{-2} & 0 & 0 \\
    0 & 0 & 0 & -\hat{\bm A}_{2} & -\hat{\bm A}_{1}  & \edit{\bm L_{1}} & -\hat{\bm A}_{-1}  & -\hat{\bm A}_{-2} & 0 \\
    0 & 0 & 0 & 0 & -\hat{\bm A}_{2} & -\hat{\bm A}_{1}  & \edit{\bm L_{2}} & -\hat{\bm A}_{-1}  & -\hat{\bm A}_{-2} \\
    0 & 0 & 0 & 0 & 0 & -\hat{\bm A}_{2} & -\hat{\bm A}_{1}  & \edit{\bm L_{3}} & -\hat{\bm A}_{-1}\\    
    0 & 0 & 0 & 0 & 0 & 0 & -\hat{\bm A}_{2} & -\hat{\bm A}_{1}  & \edit{\bm L_{4}}
\end{bmatrix} ^ {-1}.
\label{eqn:h_op1}
\end{equation}
The forcing and response vectors can be represented as
\begin{equation}
\hat{\bm f} = 
\begin{bmatrix} 
    \hat{\bm f}_{-4} \\
    \hat{\bm f}_{-3} \\
    \hat{\bm f}_{-2} \\
    \hat{\bm f}_{-1} \\
    \hat{\bm f}_{0} \\
    \hat{\bm f}_{1} \\
    \hat{\bm f}_{2} \\
    \hat{\bm f}_{3} \\
    \hat{\bm f}_{4} \\     
\end{bmatrix},
\hat{\bm q} = 
\begin{bmatrix} 
    \hat{\bm q}_{-4} \\
    \hat{\bm q}_{-3} \\
    \hat{\bm q}_{-2} \\
    \hat{\bm q}_{-1} \\
    \hat{\bm q}_{0} \\
    \hat{\bm q}_{1} \\
    \hat{\bm q}_{2} \\
    \hat{\bm q}_{3} \\
    \hat{\bm q}_{4} \\
\end{bmatrix}.
\label{eqn:forcing_harmonic}
\end{equation}

The accuracy of the obtained resolvent system in Fourier space depends on how well the chosen frequencies capture the relevant flow information, and it is affected by the truncation limit imposed on the frequency range. Although it is feasible to reduce the infinite-dimensional harmonic operator to a finite dimension based on perturbation frequencies, the choice of frequency range is consequential. To demonstrate the influence of truncation on response modes, consider a periodic flow with a dominant first harmonic, such that the base flow can be expressed as
\begin{equation}
\bar{\bm q}(t) = \hat{\bar{\bm q}}_0 + \hat{\bar{\bm q}}_1e^{\text{i}\omega t} + \hat{\bar{\bm q}}_{-1}e^{-\text{i}\omega t}.
\label{eqn:1st_term}
\end{equation}
Assume that the LNS equations are subjected to a constant forcing $\bm f(t) = \hat{\bm f}_0$. The LNS operator exhibits frequency content at 0 and $\pm \omega$, indicating that a constant forcing will directly induce a response at these frequencies. However, higher harmonics are also indirectly stimulated, and the triggering of responses at higher harmonics follows a cascade pattern, with norms gradually decaying towards zero, \ie $\lim_{j \to \infty} ||\hat{\bm q}_j|| = 0$. Hence, the truncation must be set at a sufficiently high level to preserve the response norms.

In practice, by defining $\Omega_{q}$ for both input perturbation and the response, we compute accurate harmonic resolvent modes as long as the norms of higher frequency modes become less relevant within the spectrum. In other words, if we set both forcing and response frequencies sufficiently high and, upon computing harmonic resolvent modes, find that the norms of input and output modes with higher frequency content are relatively small, it indicates convergence, and including higher frequencies has negligible impact on the results. In a case, for instance, where response modes with higher frequency content remain relatively important, extending the frequency range of the output modes is necessary without changing the input frequency content. In our algorithm, the input and output frequency ranges are adjustable as needed.

\subsection{Cross-frequency harmonic resolvent analysis} \label{sec:CrossFreq}

Harmonic resolvent analysis captures the interaction between various frequencies within $\Omega_{\bar q}$ and $\Omega_{q}$, and the typical objective is to identify the optimal forcing mode, potentially spanning various frequencies, that produces the optimal response across a range of frequencies. Alternatively, one can compute the unit-norm forcing at one frequency $\omega_1 \in \Omega_{q}$ that triggers the most amplified response at another frequency $\omega_2 \in \Omega_{q}$. This is achieved by defining
\begin{equation}
\bm H_{\omega_2, \omega_1} = \bm C \bm H \bm B
\label{eqn:CrossFreqOpt}
\end{equation}
and computing the modes and gains following the same procedure as before. Here, matrices $\bm B$ and $\bm C$ are constructed to extract the $\omega_1$ and $\omega_2$ frequencies from the forcing and response modes, respectively. This represents a special case of the modified harmonic resolvent operator in \eqref{eqn:mod_Hres_op}.

\subsection{Subharmonic resolvent analysis} \label{sec:sub}

The frequency content of the forcing inputs is typically composed of harmonics of the base flow frequency. However, \citet{PadovanRowley22} have shown that the harmonic resolvent system may exhibit sensitivity to subharmonic inputs. For instance, in the case where the fundamental base flow frequency is denoted by $\omega_f$, the linearized system may be sensitive to forcing inputs with $\gamma < \omega_f$, such as $\omega_f/2$. More generally, if the frequency content of a periodic LNS operator is limited to the frequency range $\Omega_{\bar q}$, it may be desirable to compute input-output modes with frequencies outside of this range, \ie, $\gamma \notin \Omega_{\bar q}$. In such cases, the subharmonic resolvent modes become relevant.

To identify unique subharmonic inputs, we can take advantage of the linear nature of the harmonic resolvent operator, which allows interaction with the fundamental frequency $\omega_f$ and its harmonics. It can be demonstrated that the interval $(-\omega_f/2, \omega_f/2]$ encompasses all sets of subharmonic frequencies where the input-output modes are unique. As discussed in $\S$\ref{sec:hres}, a forcing with a frequency $\gamma \in (-\omega_f/2, \omega_f/2]$ can only trigger responses at frequencies $\gamma + j\omega_f$, where $j$ is an integer. Therefore, the input and output modes exist within the frequency range $\Omega_{\gamma} = \gamma \oplus \Omega_{\bar q}$, where $\oplus$ denotes element-wise addition. Harmonics of $\gamma$ trigger a unique set of outputs as long as $j \gamma \le |\omega_f|$. Due to the linear relationship, we can show that all harmonics are isolated sets that need to be studied separately. Note that $-\omega_f/2$ is excluded since $\Omega_{-\omega_f/2} = \Omega_{\omega_f/2}$, resulting in redundancy. For $\gamma \in \Omega_{\gamma}$, one can derive the subharmonic resolvent system as \citep{PadovanRowley22}
\begin{equation}
\hat{\bm q} =  (\text{i}\gamma \bm I - \bm T)^{-1}\hat{\bm f}.
\label{eqn:T_subharmonic}
\end{equation}

To further elucidate the point, consider an example where the subharmonic frequency of interest is $\omega_f/5$ along with its harmonics. Note that the valid harmonics are the ones that fulfill the condition $j \omega_f/5 \in (-\omega_f/2, \omega_f/2]$, which includes the set $\{-2/5, -1/5, 0, 1/5, 2/5\}$. While $\Omega_{0} = 0 \oplus \{-N_b/2, \hdots, N_b/2\}$, where $N_b$ is the number of active frequencies within the base flow, is identical to harmonic inputs, the sets $\Omega_{-2/5}$, $\Omega_{-1/5}$, $\Omega_{1/5}$, and $\Omega_{2/5}$ represent distinct input-output modes. In general, for any two distinct subharmonic frequencies $\gamma_1$ and $\gamma_2$ within the interval $(-\omega_f/2, \omega_f/2]$, the corresponding subharmonic modes in $\Omega_{\gamma_1}$ are decoupled from those in $\Omega_{\gamma_2}$ as there exists no common frequency $\gamma_1 - \gamma_2 \in \Omega_{\bar q}$. While our algorithm described in $\S$\ref{sec:time-stepping} is described for harmonic resolvent analysis, it can be easily extended to compute subharmonic resolvent modes, as described in Appendix \ref{appA}.  

\edit{The approach used to eliminate the singularity of $\bm T$ described in $\S$\ref{sec:singularity} is not applicable for subharmonic resolvent analysis ($\gamma \neq 0$).  \citet{PadovanRowley22} formulated an alternative approach in which the domain of $\bm T$ is constrained to a physically meaningful subspace for all choices of $\gamma$. In this approach, an oblique projection operator is defined as
\begin{equation}
\bm P(t) =  \bm I - \bm v(t) \bm u^*(t),
\label{eqn:Poblique}
\end{equation}
where $\bm u$ and $\bm v$ are defined the same as in $\S$\ref{sec:singularity}. The restricted operator can be written as 
\begin{equation}
\bm T_p =  \bm T \hat{\bm P}.
\label{eqn:Toblique}
\end{equation}
The resulting range of $\bm{T}_p$ is deliberately designed to be invariant under $T$. This projection facilitates steps within the RSVD-LU algorithm, effectively addressing singularities. For more details on oblique projection properties and the efficient computation of $\bm{u}$ and $\bm{v}$, we refer the reader to \citet{PadovanRowley22}.
}

\subsection{Computing harmonic resolvent modes using RSVD-LU} \label{sec:RSVD}

The RSVD algorithm \citep{Halkoetal11} is a randomized linear algebra technique developed to efficiently identify the singular vectors with the highest gains in a given matrix. In the context of harmonic resolvent analysis, the matrix of interest is the harmonic resolvent operator. This operator, similar to the resolvent operator, relies on the inverse of a sparse matrix, as shown in \eqref{eqn:Hres}, making the computation of modes challenging. Modifying the original RSVD algorithm is necessary to address these challenges, enabling harmonic resolvent analysis. The modifications for resolvent analysis have been extensively documented in the literature by \citet{Ribeiroetal20} and \citet{Farghadanetal23}, while the outline for the harmonic resolvent operator can be found in the appendix of \citet{Padovanetal20}. 

In brief, the computation involves computing the action of $\bm H$ and $\bm H^*$ for sketching the range and image of the harmonic resolvent operator, respectively, and approximating the leading harmonic modes. Computing actions of $\bm H$ and $\bm H^*$, however, requires solving linear systems
\begin{equation}
\begin{gathered}
\bm T\hat{\bm q} = \hat{\bm f},
\\
\bm T^*\hat{\bm q} = \hat{\bm f},
\label{eqn:action}
\end{gathered}
\end{equation}
respectively, in Fourier space. These linear systems are solved via LU-decomposition of $\bm T$, but due to its large size and ill-conditioned nature, computing its LU decomposition can be a formidable computational obstacle, particularly for flows with three inhomogeneous directions. We propose an alternative way in the following section.

\section{Computing harmonic resolvent modes using time stepping} \label{sec:time-stepping}

In this section, we show how time stepping can be used to efficiently compute harmonic resolvent modes. The same concept was first used by~\citet{Monokrousosetal10} in the context of the resolvent analysis and further optimized by \citet{Martinietal21} and \citet{Farghadanetal23}. Here, we introduce the extension of our algorithm to compute the harmonic resolvent modes and gains.

\subsection{Computing the action of \texorpdfstring{$\bm H$}{H} using time stepping} \label{sec:H_time-stepping}

\begin{figure}
\centering
\includegraphics[width=\textwidth]{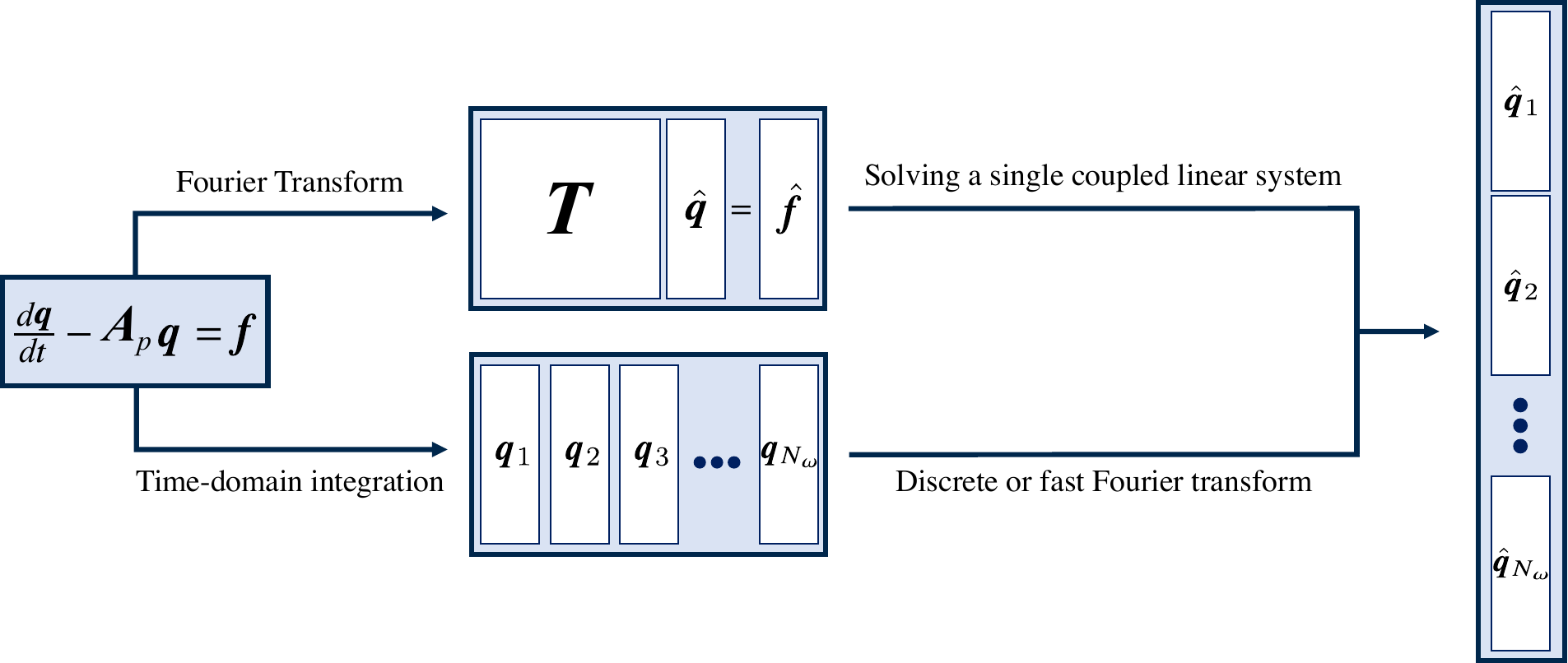}
\caption{Flowchart depicting the action of $\bm H$ on a forcing comprised of $N_{\omega}$ frequencies within the RSVD-LU (top route) and the RSVD-\Dt (bottom route) algorithms. Both routes produce the same result, but the bottom route is computationally advantageous for large systems.}
\label{fig:flowchart}
\end{figure}

Computing the action of $\bm H$ in the Fourier space poses a bottleneck within the RSVD-LU algorithm. To overcome this limitation, we propose to use time-stepping as an effective surrogate approach. Given that $\Omega_{\bar q}$ represents the frequency content of the base flow around which $\bm T$ is constructed, we define the time-domain linearized operator as
\begin{equation}
\bm A_p(t) = \sum_{j\omega_f \in \Omega_{\bar q}} \hat{\bm A}_{p, j} e^{\text{i} j\omega_f t}.
\label{eqn:Ap}
\end{equation}
The action of $\bm H$ on $\hat{\bm f}$ in Fourier space is expressed in \eqref{eqn:Hres_eqn}, where the frequency content of the forcing lies within $\Omega_{q}$. 

The steady-state solution of \eqref{eqn:linsys_harmonic} is given by
\begin{equation}
\hat{\bm q}_s = \bm H \hat{\bm f},
\label{eqn:direct_action}
\end{equation}
when subjected to the forcing represented as
\begin{equation}
\bm f = \sum_{j\omega_f \in \Omega_{q}} \hat{\bm f}_{j}e^{\text{i} j\omega_f t}. 
\label{eqn:harmonic_series}
\end{equation}
The steady-state response can be obtained through time-stepping before taking a Fourier transform. The time-domain forcing is constructed to include frequencies identical to those present in $\hat{\bm f}$. Since the time-domain and frequency-domain frequencies are the same, both approaches yield identical results, \ie $\hat{\bm q}_s = \hat{\bm q}$. In practice, the time-stepping method is inherently discrete, and the frequencies are retained up to the maximum limit determined by the chosen time step (below the Nyquist frequency \citep{Nyquist28}). By employing a suitably small time step, we can capture frequencies up to the desired limit.

The equivalence between computing the action of $\bm H$ in both the RSVD-LU and RSVD-$\Delta t$ algorithms is depicted in figure \ref{fig:flowchart}. In the upper route, within the RSVD-LU algorithm, the LNS equations are first transformed into Fourier space before solving a coupled linear system in the frequency domain for a given forcing. In the lower route, within the RSVD-$\Delta t$ algorithm, the LNS equations are initially integrated in the time domain before being transformed into Fourier space. Both routes produce identical outputs so long as numerical artifacts are minimized.

\subsection{RSVD-\texorpdfstring{$\Delta t$}{deltat} for harmonic resolvent analysis} \label{sec:RSVDt}

By recognizing the equivalence between time stepping and solving linear systems in the frequency domain, we can effectively address the limitations of the RSVD-LU algorithm, leading to the development of RSVD-$\Delta t$. \edit{The time-stepping aspect of RSVD-\Dt relies on the linearized operator $\bm A_p$ without explicitly constructing $\bm H$ in Fourier space.} The algorithmic steps for extending RSVD-\Dt to harmonic systems are presented in Algorithm~\ref{alg:alg_RSVDt}.


\begin{algorithm} 
\caption{RSVD-\Dt for harmonic resolvent analysis}
\label{alg:alg_RSVDt}
\begin{algorithmic}[1]

\State \textbf{Inputs:} $\bm A_p, k, q, \Omega_{q},\Omega_{\bar q}, $TSS,\,$ dt, T_t$
    
    \State $\hat{\boldsymbol{\varTheta}} \leftarrow \;$randn$(NN_{\omega}, k)$ 
    \label{algt:stp1}
    \Comment{Create random test matrices} 
    
    \State $\hat{\bm Y}  \leftarrow \;$\texttt{DirectAction}$(\bm A_p,\hat{\boldsymbol{\varTheta}}, $TSS,\,$dt, T_t)$ \label{algt:stp2}
    \Comment{Sample the range of $\bm H$}  
        
    \If{$q > 0$}  \Comment{Optional power iteration} \label{algt:stp3_1}
    \State $\hat{\bm Y} \leftarrow \;  $\texttt{PI}$(\bm A_p, \hat{\bm Y}, q, $TSS,\,$dt, T_t)$ 
    \Comment{Power iteration with time-stepping}
    \label{algt:stp3_2}
    
    \EndIf
    
        
    \State $\hat{\bm Q}_{\varOmega} \leftarrow \;$qr$(\hat{\bm Y}_{\varOmega})$ 
    \label{algt:stp4}
    \Comment{Build the orthonormal subspace $\hat{\bm Q}_{\varOmega}$}

        
    \State $\hat{\bm S}  \leftarrow \;$ \texttt{AdjointAction}$(\bm A_p^*,\hat{\bm Q}, $TSS,\,$dt, T_t)$  \label{algt:stp5}
    \Comment{Sample the image of $\bm H$}

        
    \State $(\tilde{\bm U}_H , \boldsymbol{\varSigma}_H , \bm V_H) \leftarrow \;$svd$(\hat{\bm S}_H)$  \label{algt:stp6}
    \Comment{Obtain $\boldsymbol{\varSigma}_H , \bm V_H$}
    
    \State $\bm U_H  \leftarrow \; \hat{\bm Q}_{\varOmega} \tilde{\bm U}_H$
    \Comment{Recover $\bm U_H$} \label{algt:stp7}

\State \textbf{Outputs:} $\bm U_H , \boldsymbol{\varSigma}_H , \bm V_H$

\NoNumber{\footnotesize{Algorithm 1. \edit{Inputs: linearized operator $\bm A_p$, number of modes $k$, number of power iterations $q$, frequency range of the perturbations and LNS operator $\Omega_{q}$ and $\Omega_{\bar q}$, respectively, time-stepping scheme abbreviated as TSS (\eg backward Euler), time step $dt$, and the transient length $T_t$. Outputs: $k$ response modes $\bm U_H$, $k$ forcing modes $\bm V_H$ and the corresponding gains $\boldsymbol{\varSigma}_H$.} Here, $k, q, \Omega_{q},\Omega_{\bar q}$ are common parameters with RSVD-LU. $(\cdot)_{\varOmega}$ indicates all frequencies are merged into a single column. \texttt{DirectAction} and \texttt{AdjointAction} are functions that solve the direct and adjoint LNS equations, respectively, with a predefined forcing. \edit{\texttt{PI} is a function that performs the power iteration.}}}

\end{algorithmic}
\end{algorithm}


The algorithm proposed in this study is based on the RSVD-\Dt algorithm introduced by \citet{Farghadanetal23}, and we will provide a concise overview of the key steps. Initially, a random matrix $\hat{\boldsymbol{\varTheta}} \edit{\in \mathbb C^{NN_{\omega}\times k}}$ is generated (line \ref{algt:stp1}) to compute the sketch of $\bm H$ through time stepping (line \ref{algt:stp2}). \edit{The dimensions of the random matrix are determined by the state dimension $N$, the number of frequencies to resolve $N_{\omega}$, and the desired number of modes to compute $k$.}. \texttt{DirectAction} is a function that computes the action of $\bm H$ onto a given forcing using time-stepping and transforms the steady-state responses to Fourier space. \edit{As outlined in $\S$\ref{sec:H_time-stepping}, \texttt{DirectAction} solves \eqref{eqn:linsys_harmonic} with zero initial condition over a sufficiently long time interval, causing the initial transient response to dissipate. $T_t$ represents the duration of this interval before the transient diminishes. The forcing term in \eqref{eqn:linsys_harmonic} is constructed via inverse Fourier transform of $\hat{\boldsymbol{\varTheta}}$ and the steady-state response undergoes a Fourier transform to obtain $\hat{\bm Y}$.} Next, an optional power iteration \edit{(\texttt{PI})} is performed (lines \ref{algt:stp3_1} and \ref{algt:stp3_2}) via $q$ successive applications of \texttt{DirectAction} and \texttt{AdjointAction}, enhancing the accuracy of harmonic resolvent modes. QR decomposition is performed to obtain $\hat{\bm Q}_{\varOmega}$ (line \ref{algt:stp4}) before constructing $\bm S$ via time stepping (line \ref{algt:stp5}). \texttt{AdjointAction} is the function that computes the action of $\bm H^*$ onto a given forcing using time-stepping and transforms the steady-state responses to Fourier space. \edit{This function is similar to \texttt{DirectAction}, but it evolves the adjoint system. The forcing term in \texttt{AdjointAction} is constructed via the inverse Fourier transform of $\hat{\bm Q}$.} An SVD (line \ref{algt:stp6}) is conducted to obtain the optimal forcing modes $\bm V_H \in \mathbb C^{NN_{\omega}\times k}$ and gains $\boldsymbol{\varSigma}_H \in \mathbb R^{k\times k}$ of $\bm H$. Lastly, the optimal response modes $\bm U_H \in \mathbb C^{NN_{\omega}\times k}$ are recovered in line \ref{algt:stp7}.

Compared to the RSVD-$\Delta t$ algorithm for resolvent analysis \cite{Farghadanetal23}, two notable differences are present. In the extension of RSVD-$\Delta t$, the response modes at various frequencies are merged prior to both the QR decomposition and the SVD steps. This deviation from the original algorithm is motivated by the nature of the problem that requires interactions between frequencies, unlike in resolvent analysis where the QR decomposition and SVD are performed separately for each individual frequency of interest. The second difference is the generation of LNS operators over time, whereas in resolvent analysis, the LNS operator remains constant throughout the integration.

If $\bm T$ approaches machine precision singularity, it is necessary to project out the forcing along the $\bm u$ direction before the \texttt{DirectAction} and along $\bm v$ before the \texttt{AdjointAction}, as described in $\S$\ref{sec:singularity}. Moreover, to mitigate numerical artifacts, it is advisable to project out the response along $\bm v$ and $\bm u$ after the \texttt{DirectAction} and \texttt{AdjointAction}, respectively.

\section{Computational complexity} \label{sec:computation}

A brief comparison is conducted between memory requirements and CPU cost of the RSVD-LU and RSVD-\Dt algorithms. We also propose strategies to minimize the memory consumption. \edit{Throughout, we assume that the linearized operator $\bm A_p$ is sparse, \ie that its number of non-zero entries scales as $\texttt{nnz}(\bm A_p) \sim O(N)$.  Sparse operators are obtained when using sparse discretization schemes such as finite differences, finite volume, or finite elements.} For more detailed information on the CPU and memory scaling of the LU decomposition of the resolvent operator, we refer readers to \citet{Farghadanetal23}. To offer empirical insights and confirmation of the theoretical scalings discussed below, we present a two-dimensional test case in $\S$\ref{sec:TestCases}.

\subsection{Memory benefits of RSVD-\texorpdfstring{$\Delta t$}{deltat}} \label{sec:mem}

In the case of resolvent analysis, the memory scaling of the LU decomposition is empirically $O(N^{1.2})$ and $O(N^{1.6})$ for two- and three-dimensional systems, respectively \citep{Towneetal22, Farghadanetal23}. The memory scaling of LU decomposition is expected to be worse for the harmonic resolvent operator, as the lower- and upper-triangular matrices are denser. When $\bm T$ is singular or nearly singular, the computational task becomes even more challenging, as discussed in~\citet{Padovanetal20}. \edit{An alternative approach for solving \eqref{eqn:action} involves leveraging iterative solvers, such as Krylov subspace methods \citep{Rigasetal21, Wuetal22}. This approach eliminates the need for computing the LU decomposition and becomes more memory-efficient, especially for large systems. However, finding a suitable preconditioner remains a challenging task.}

On the other hand, the memory usage of RSVD-\Dt for harmonic resolvent analysis is contingent on the sizes of various matrices, specifically the sparse LNS operators $\hat{\bm A}_p \in \mathbb{C}^{N\times N\times N_b}$, and the dense forcing matrix $\hat{\bm F} \in \mathbb{C}^{NN_{\omega}\times k}$ and response matrix $\hat{\bm Q} \in \mathbb{C}^{NN_{\omega}\times k}$, all of which are stored in Fourier space. The scaling for resolvent analysis is $O(N N_\omega)$, and the sole extra space is for the storage of LNS operators.

\begin{figure}[h]
\centering
\includegraphics[width=\textwidth,height=\textheight,keepaspectratio]{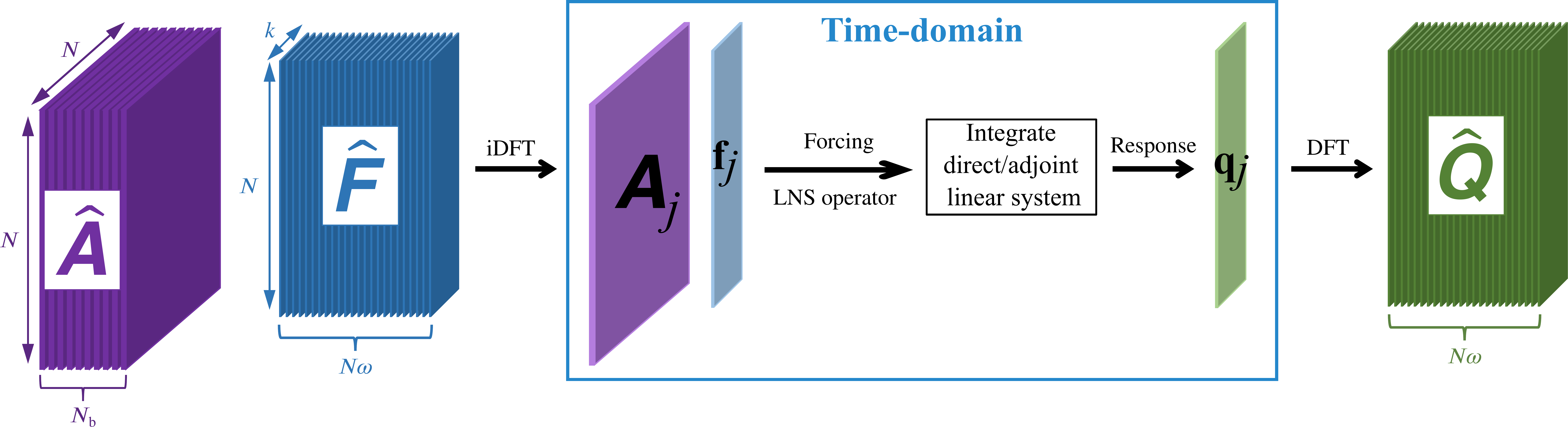}
\caption{Schematic of the action of $\bm H$ with streaming DFT/iDFT to transform between the Fourier and time domains.}
\label{fig:dir_adj_scheme}
\end{figure}

One approach to generating LNS operators for all time points involves retaining $N_b$ coefficients in Fourier space and subsequently constructing $\bm A_{\varOmega} = \{\bm A_{p,1},\bm A_{p,2},\bm A_{p,3},\dots,\bm A_{p,N_t}\}$ once for all time steps within a period. However, this approach can be memory-intensive for large systems, especially when $N_t \sim O(10^3-10^5)$. An alternative strategy to reduce memory consumption is to create LNS operators on the fly, as shown in figure \ref{fig:dir_adj_scheme}. As a result, no time-domain LNS operator is permanently stored in memory. Each LNS operator is created on the fly using the discrete Fourier transform (DFT) of $N_b$ Fourier coefficients,
\begin{equation}
\bm A_p(t) = \sum_{j = -N_b/2}^{N_b/2} \hat{\bm A}_{p, j} e^{\text{i} j\omega_f t}.
\label{eqn:periodic_A_Nw}
\end{equation}
Since $\bm A_p$ is very sparse, $\texttt{nnz}(\bm A_p) \sim O(N)$, resulting in an overall memory consumption of $O(NN_b)$. The methods employed for generating LNS operators, both in the streaming and memory-intensive approaches, mirror those used for creating forcing terms in resolvent analysis, as elaborated in \citet{Farghadanetal23}. Additionally, the CPU cost of this procedure scales as $O(\texttt{nnz}(\bm A_p)N_b)$ or $O(NN_b)$.

Streaming DFT is indeed effective in reducing memory consumption; nevertheless, additional memory savings can be specifically achieved for real-valued LNS operators. In many cases, $\bm A_p$ is real-valued, \ie confined to $\mathbb R^{N\times N}$, yielding $\hat{\bm A}_{p, j}$ = $\bar{\hat{\bm A}}_{p, -j}$, where $\bar{(\cdot)}$ denotes the complex conjugate. Hence, $\hat{\bm A}_{p, j}$ with $j \ge 0$ suffices to construct $\bm A_p$ as
\begin{equation}
\bm A_p(t) = \hat{\bm A}_{p, 0} + 2\mathcal Re\left (\sum_{j = 1}^{N_b/2} \hat{\bm A}_{p, j} e^{\text{i} j\omega_f t} \right),
\label{eqn:realA}
\end{equation}
resulting in halving the number of Fourier coefficients of the LNS operator to $\lfloor N_b/2 \rfloor + 1$. Here, $\mathcal Re$ represents the real part.

Another opportunity to reduce memory usage for real-valued LNS matrices arises from the symmetry between positive and negative frequencies. Rewriting \eqref{eqn:inverse_harmonic_res} for $-k\omega_f$ yields 
\begin{equation}
\text{i}(-k\omega_f) \hat{\bm q}_{-k} - \sum_{j = -\infty}^{\infty} \hat{\bm A}_{-k + j} \hat{\bm q}_{-j} = \hat{\bm f}_{-k},
\label{eqn:NegativeT}
\end{equation}
and the complex conjugate version of \eqref{eqn:inverse_harmonic_res} becomes
\begin{equation}
-\text{i}k\omega_f \bar{\hat{\bm q}}_k - \sum_{j = -\infty}^{\infty} \bar{\hat{\bm A}}_{k - j} \bar{\hat{\bm q}}_j = \bar{\hat{\bm f}}_k.
\label{eqn:ConjT}
\end{equation}
Both equations \eqref{eqn:NegativeT} and \eqref{eqn:ConjT} yield the same output for a given $\bar{\hat{\bm f}}_k = \hat{\bm f}_{-k}$ since $\bar{\hat{\bm A}}_{k - j} = \hat{\bm A}_{-k + j}$, inducing the harmonic resolvent response and forcing modes containing $\bm U_{H, j} = \bar{\bm U}_{H, -j}$ and $\bm V_{H, j} = \bar{\bm V}_{H, -j}$ for $j \ne 0$, respectively. Hence, we only keep $\lfloor N_{\omega}/2 \rfloor + 1$ Fourier coefficients of the forcing and response modes, reducing memory consumption by half when storing these dense matrices.

\subsection{CPU benefits of RSVD-\texorpdfstring{$\Delta t$}{deltat}}

The computational cost of computing harmonic resolvent modes using RSVD-LU is predominantly determined by the LU decomposition of $\bm T$. In the case of resolvent analysis, the CPU scaling of the LU decomposition is $O(N^{1.5})$ and $O(N^2)$ for two- and three-dimensional systems, respectively \citep{Towneetal22, Farghadanetal23}. When, in general, block matrix $\bm T$ contains  $\hat{\bm A}_{p, i}$ terms, and yields a denser matrix than block diagonal with a complex sparsity pattern, the scaling is expected to be $O((N_{\omega}N)^{1.5})$ or $O((N_{\omega}N)^2)$ or worse. \edit{Alternatively, iterative solvers may outperform LU decomposition in terms of CPU scalability, contingent on improving the condition number of $\bm T$ via applying an appropriate preconditioner \citep{Rigasetal21, Wuetal22}, which itself may or may not be inexpensive to obtain.}

The cost of using RSVD-\Dt is directly tied to the cost of computing the actions of $\bm H$ and $\bm H^*$. In resolvent analysis, the total CPU cost is proportional to the state dimension $N$. This cost can be broken down into three main components: $(i)$ time-stepping, \eg Adams-Bashforth methods, which scales as $O(N)$ when $\bm A$ is sparse, $(ii)$ creating forcing from Fourier space to the time domain, which scales as $O(NN_tN_{\omega})$, $(iii)$ taking the response from the time domain to Fourier space, which scales as $O(NN_{\omega}^2)$. Among these components, the time-stepping part is the most significant contributor to the overall cost. The CPU cost of harmonic resolvent analysis can be assessed similarly to resolvent analysis, with one exception. For periodic flows, the LNS operator varies over time, whereas it remains constant for statistically stationary flows. We showed in $\S$\ref{sec:mem} that the cost of creating LNS operators over time scales as $O(NN_b)$ for sparse matrices.

In summary, the CPU and memory costs of RSVD-\Dt exhibit linear scaling with respect to the state dimension $N$, regardless of whether the base flow is steady or periodic. Moreover, the dominant terms in the CPU cost are independent of the number of retained frequencies $N_{\omega}$. Both of these properties make the RSVD-\Dt algorithm considerably more scalable than RSVD-LU for harmonic resolvent analysis.   
 
\section{Minimizing the CPU cost of RSVD-\texorpdfstring{$\Delta t$}{deltat}} \label{sec:Opt}

Similar to time stepping in the context of resolvent analysis, periodic systems may also experience a slow decay of transient response, which is undesirable, since we require the steady-state response in isolation to compute the action of the harmonic resolvent operator using time stepping. This phenomenon might be linked to instances when $\bm T$ is near singularity. Regardless, in cases where the LNS equations are absolutely stable but exhibit slowly decaying modes, we propose a strategy that can effectively reduce the CPU cost of time stepping.

\subsection{Stability analysis: Floquet theorem and transient response} \label{sec:stability}

The steady-state response of the LNS equations, subject to forcing as described in equation \eqref{eqn:linsys_harmonic}, is of interest for our analysis. In the case of a linear time-invariant system where $\bm A$ is independent of time, the decay rate is controlled by the least-damped eigenvalue of $\bm A$. This decay rate can be estimated by integrating a homogeneous system over a sufficiently long period. However, for a linear time-periodic system, the decay rate is determined by the least-damped Floquet exponent.

Assuming $\boldsymbol{\Phi}(t)$ is the fundamental solution of \eqref{eqn:linsys_harmonic}, the monodromy matrix is constructed as 
\begin{equation}
\bm M = \boldsymbol{\Phi}(0)^{-1}\boldsymbol{\Phi}(T),
\label{eqn:monodromy}
\end{equation}
where $T$ is the period of the flow. When $\boldsymbol{\Phi}(0) = \bm I, \boldsymbol{\Phi}(t)$ is referred to as the principal fundamental matrix, and the monodromy matrix simplifies to $\boldsymbol{\Phi}(T)$. In other words, we evaluate the principal fundamental matrix at the end of the first period. The eigenvalues of $\bm M$, known as Floquet multipliers $\mu_j$, allow us to determine the Floquet exponents $\lambda_j = \log(\mu_j)/T$. The least-damped Floquet exponent has the smallest real part and it determines the decay rate of the transient response. If any mode is located in the unstable half-plane, \ie $\mathcal Re(\lambda_j) > 0$, the system is globally unstable. 

It is worth noting that the transient length remains independent of the steady-state period. This characteristic is particularly important as it ensures that the length of integration does not rely on $T$, thereby avoiding potential deterioration in the performance of RSVD-\Dt for flows with low periodicity. In practice, both finding the principal fundamental matrix and determining the Floquet exponents can be computationally expensive, especially for large-scale systems. As with autonomous cases, it is possible to run a homogeneous simulation to analyze the long-term behavior of the transient response for periodic flows. This equivalence has been illustrated on a test case in $\S$\ref{sec:TestCases}.

If $\bm T$ is singular but all other modes are stable, then the Floquet exponent with the smallest real part will be zero with $\widehat{d\bm{\bar q}/dt}$ as the corresponding Floquet mode. \edit{This is a well-known fact from Floquet theory \citep{Wereley90}} and can also be seen from \eqref{eqn:NS_timeDerivative}, where the phase shift direction is a non-trivial solution of the homogenous system. \citet{LeclercqSipp23} have also shown that $d\bm{\bar q}/dt$ is associated with a zero Floquet exponent under the assumption that the base flow satisfies the Navier-Stokes equations as in \eqref{eqn:NS}.

\subsection{Efficient transient removal}

\par Our strategy leverages the periodic nature of the steady-state part and the linear evolution of the transient part of the solution to directly compute and eliminate the undesired transient portion. For a given pair of solutions $\bm q_1$ and $\bm q_2$ at times $t_1$ and $t_2 = t_1 + T$, respectively, they can be expressed as a sum of their steady-state and transient components as
\begin{equation}
\begin{aligned}
\bm q_1 &= \bm q_{s, 1} + \bm q_{t, 1},
\\
\bm q_2 &= \bm q_{s, 2} + \bm q_{t, 2},
\label{eqn:1}
\end{aligned}
\end{equation}
where $\bm q_{t, 1}$ and $\bm q_{t, 2}$ represent the transient part which decays to zero as $t \to \infty$ and $\bm q_{s, 1}$ and $\bm q_{s, 2}$ denote the steady-state part which evolves periodically. The time distance between two snapshots is one period, hence
\begin{equation}
\bm q_{s, 2} = \bm q_{s, 1}.
\label{eqn:2}
\end{equation}
Also, the evolution of the transient part can be expressed as
\begin{equation}
\bm q_{t, 2} = \boldsymbol{\Phi}(T)\bm q_{t, 1},
\label{eqn:3}
\end{equation}
where $\boldsymbol{\Phi}$ is the principal fundamental matrix of \eqref{eqn:linsys_harmonic}. Simplifying \eqref{eqn:1}, \eqref{eqn:2}, and \eqref{eqn:3} leads to 
\begin{equation}
(\boldsymbol{\Phi}(T) - \bm I)\bm q_{t, 1} = \bm b,
\label{eqn:4}
\end{equation}
where $\bm b = \bm q_2 - \bm q_1$ is known.

By obtaining $\bm q_{t, 1}$ from solving \eqref{eqn:4}, the steady-state solution can be recovered as $\bm q_{s,1} = \bm q_1 - \bm q_{t,1}$. The central challenge in removing the transient is the computational cost associated with solving the linear system in \eqref{eqn:4}. Instead, we employ Petrov-Galerkin (or Galerkin) projection to obtain an approximate solution to \eqref{eqn:4} in a more cost-effective manner.

A low-dimensional representation of the transient part can be expressed as 
\begin{equation}
\bm q_{t, 1} = \boldsymbol{\phi} \boldsymbol{\beta}_1,
\label{eqn:5}
\end{equation}
where $\boldsymbol{\phi} \in \mathbb{C}^{N \times r}$ is an orthonormal low-dimensional \edit{trial} basis with $r \ll N$, and $\boldsymbol{\beta}_1 \in \mathbb{C}^{r}$ represents the coefficients describing the transient response in this basis. Substituting \eqref{eqn:5} into \eqref{eqn:4}, the linear system 
\begin{equation}
(\boldsymbol{\Phi}(T) - \bm I)\boldsymbol{\phi} \boldsymbol{\beta}_1 = \bm b
\label{eqn:6}
\end{equation}
is overdetermined. We use Petrov-Galerkin projection with a low-dimensional \edit{test} basis $\boldsymbol{\psi} \in \mathbb{C}^{N \times r}$ to close \eqref{eqn:6}, yielding
\begin{equation}
\tilde{\bm M} \boldsymbol{\beta}_1 = \boldsymbol{\psi}^* \bm b,
\label{eqn:7}
\end{equation}
where 
\begin{equation}
\tilde{\bm M} = \boldsymbol{\psi}^*(\boldsymbol{\Phi}(T) - \bm I)\boldsymbol{\phi}
\label{eqn:8}
\end{equation}
is the map between coefficients. Solving \eqref{eqn:7} for $\boldsymbol{\beta}_1$ is inexpensive due to its reduced dimension, and from \eqref{eqn:5} the transient estimate is obtained as
\begin{equation}
\bm q_{t, 1} = \boldsymbol{\phi} (\boldsymbol{\psi}^*\boldsymbol{\phi} - \tilde{\bm M})^{-1} \boldsymbol{\psi}^*\bm b.
\label{eqn:9}
\end{equation}

The reduced operator $\tilde{\bm M}$ is obtained by integrating the columns of $\boldsymbol{\phi}$ for one period, giving $\boldsymbol{\Phi}(T) \boldsymbol{\phi}$, and then projecting $(\boldsymbol{\Phi}(T) - \bm I)\boldsymbol{\phi}$ against the columns of $\boldsymbol{\psi}$. This integration happens once and its CPU cost is equivalent to integrating the LNS equations for $r$ periods. The analogous process occurs for the adjoint equations.

\begin{figure}
\centering
\includegraphics[width=\textwidth]{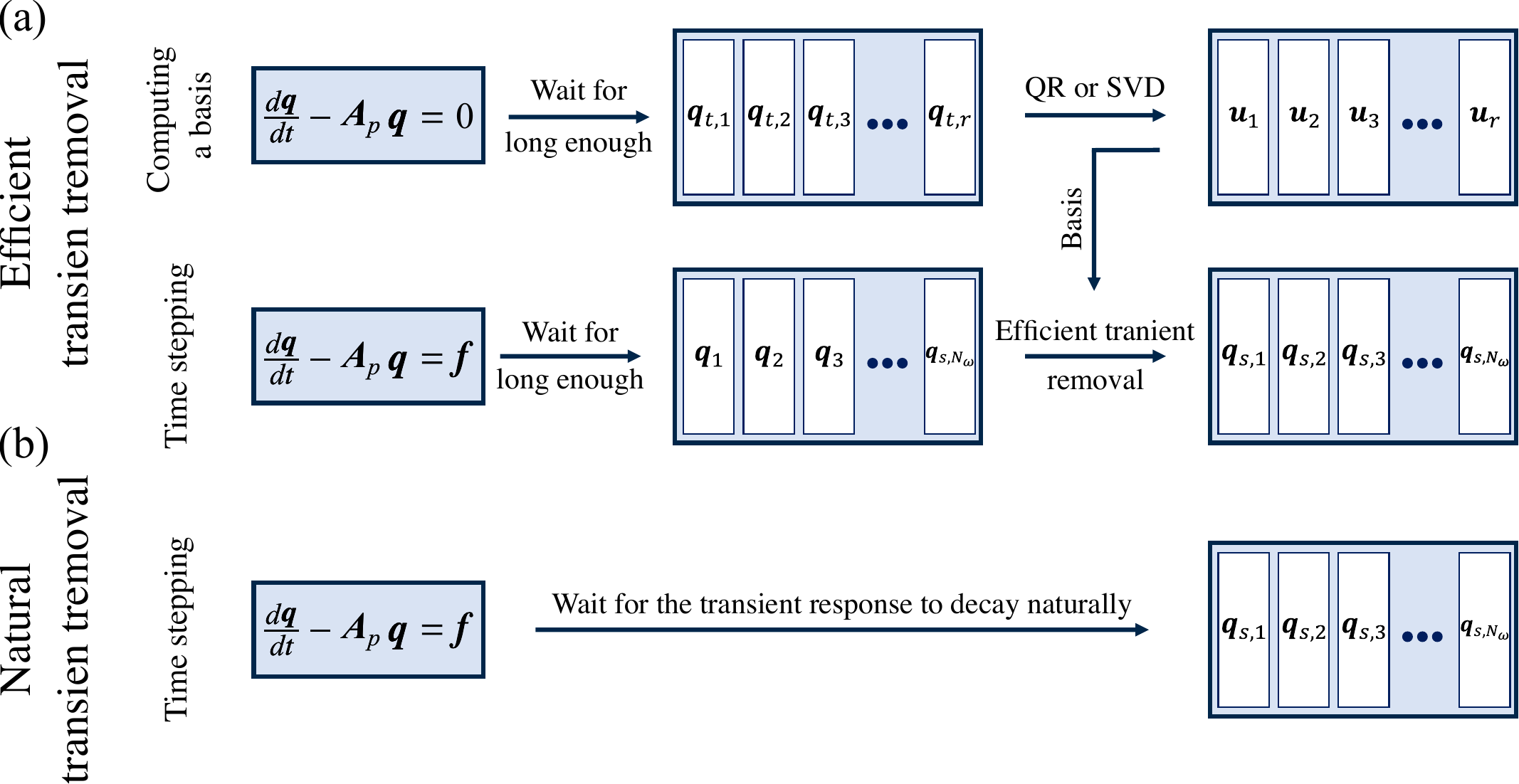}
\caption{Flowchart depicting the action of $\bm H$ using time stepping (a) with efficient transient removal strategy and (b) with natural transient decay.}
\label{fig:T_flowchart}
\end{figure}

How should the trial and test bases be selected? We have found a \edit{trial} basis spanning the least-damped modes to be effective. Rather than determining these modes through an exhaustive Floquet analysis, we employ a homogenous simulation as illustrated in the top row of the flowchart in figure \ref{fig:T_flowchart}. By initiating a sufficiently long simulation from a normalized random initial condition, the system asymptotically converges to the least-damped modes. The longer the integration length, the lower-dimensional the space becomes. As demonstrated in a test case in $\S$\ref{sec:TestCases}, the dimension of the \edit{trial} bases can, in some cases, be as small as a single column. Our test cases employ Galerkin projection with identical trial and test bases.

The obtained \edit{trial} basis effectively captures the transient response at a specific phase $\theta$. Utilizing $\boldsymbol{\phi}$ and performing Galerkin projection, we obtain $\tilde{\bm M}$. The time stepping is then carried out for a sufficiently long duration to allow the initial transients to vanish before applying \eqref{eqn:7} to determine the transient at phase $\theta$ as shown in the middle row of the flowchart in figure \ref{fig:T_flowchart}. Once the steady-state response at the same phase is updated as $\bm q_{s,1} = \bm q_1 - \bm q_{t,1}$, it is used as a new initial condition, which is synchronized with the forcing and will not excite a transient response (within the span of the bases used to construct $\tilde{\bm M}$). Then, integration for one period is sufficient, and $N_{\omega}$ steady-state snapshots are collected as desired. The resulting steady-state snapshots are identical to those obtained by a prolonged wait in the case of natural decay, as illustrated in the bottom row of the flowchart in figure \ref{fig:T_flowchart}. This procedure is also applicable to the adjoint LNS equations. Given that the adjoint equations differ from the LNS equations, a new basis needs to be constructed for them. Implementing this strategy can reduce the integration length by a factor of 10 or more, depending on the desired accuracy.

\section{Test cases} \label{sec:TestCases}

We assess the effectiveness of RSVD-\Dt in periodic systems using two test cases. First, we use a modified Ginzburg-Landau equation to validate the RSVD-\Dt algorithm and to illustrate the transient removal strategy. Second, we consider a periodic flow around an airfoil, similar to a test case from \citet{Padovanetal20}, to evaluate the accuracy and cost-savings of the RSVD-\Dt algorithm compared to RSVD-LU.
 
\subsection{Periodically varying complex Ginzburg-Landau equation} \label{sec:GL}

The one-dimensional complex Ginzburg-Landau equations is a widely used model for understanding and controlling the non-modal growth of instabilities in transitional and turbulent shear flows \citep{Chomazetal88, HuntCrighton91, Bagherietal09, ChenRowley11, Cavalierietal19}. Here, a modified system is used as an inexpensive test case to validate our algorithm. The original complex Ginzburg-Landau system follows the form of \eqref{eqn:linsys} with
\begin{equation}
\begin{gathered}
\mathcal A = -\nu \frac{\partial}{\partial x} + \gamma \frac{\partial^2}{\partial x^2}+ \mu(x), 
\\
\mu(x) = (\mu_0 - c_{\mu}^2) + \frac{\mu_2}{2} x^2,
\\
\bm B = \bm C = \bm I.
\label{eqn:GL_obj}
\end{gathered}
\end{equation} 
Following \citet{Bagherietal09}, we set $c_{\mu} = 0.2, \mu_2 = -0.01, \gamma = 1 - \text{i},$ and $\nu = 2 + 0.2\text{i}$. This system is globally stable when $\mu_0 < \mu_{0,cr} \approx 0.3977$ \citep{Bagherietal09}. By substituting $\mu_0(t) = \bar{\mu}_0 + \mu_p sin(\omega_f t -\pi/2)$ in place of $\mu_0$, we transform the Ginzburg-Landau equations to a periodically varying system with fundamental frequency $\omega_f$, following the form of \eqref{eqn:linsys_harmonic}. We set $\mu_0 = 0.395$, $\mu_p = 0.1$, and $\omega_f = 0.1$; the mean value of $\mu(t)$ is equal to $\mu_0$, whose value is close to the critical value $\mu_{0,cr}$ to promote significant growth. The periodic linearized operators $\bm A_{p, j}$ are constructed using a central finite-difference method for $x \in [-50, 50]$ with $N = 1000$ grid points, and we set $\bm W = \bm I$ on account of the uniform grid.

\begin{figure}[h]
\centering
\includegraphics[width=\textwidth,height=\textheight,keepaspectratio]{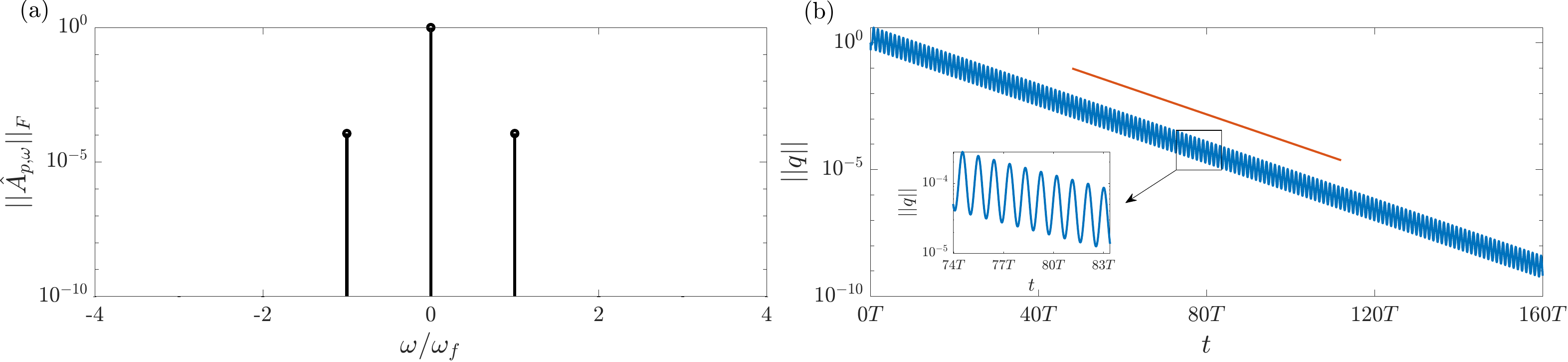}
\caption{Ginzburg-Landau test case: (a) the normalized Frobenius norm of $\hat{\bm A}_{p, \omega}$ up to the fourth harmonic. (b) The norm of the transient response over time (blue) along with the expected decay rate from Floquet analysis (red).}
\label{fig:Anorm_transient}
\end{figure}

Nine operators are built within the interval $[0, T)$, with $T = 2\pi/\omega_0 \approx 62.83$, such that $\bm A_p(t) = \bm A_p(t + T)$, before conducting a DFT to obtain $\hat{\bm A}_p$. Figure \ref{fig:Anorm_transient}(a) depicts the Frobenius norm of $\hat{\bm A}_p$ up to the fourth harmonic, normalized relative to the maximum norm. The spectrum reveals that only the mean and the first harmonic are active, as expected, since $\mu_0(t)$ oscillates at a single frequency. Accordingly, $\bm A_p$ can be reconstructed at any given time using $\hat{\bm A}_{p, 0}$, and $\hat{\bm A}_{p, \pm1}$, and the harmonic resolvent operator is constructed around $\Omega_{\bar q} = \{-1, 0, 1\}\omega_f,$ \ie $N_b = 3$. Notably, the first harmonic's norm is several orders of magnitude smaller than that of the base flow. Nonetheless, it should not be disregarded, as we will elucidate its significance in subsequent discussions.

An essential initial step involves ensuring the stability of $\bm A_p$. Typically, this is achieved by integrating the homogeneous system, \ie integrating \eqref{eqn:linsys_harmonic} with zero forcing, starting from a normalized random initial condition. However, the compact size of this system allows us to validate the time-stepping approach for computing the least-damped decaying mode against a Floquet stability analysis. The least-damped Floquet exponent, computed as $\lambda_l = -0.0021 - 0.0477\text{i}$, is expected to govern the decay rate of the transient response. Figure \ref{fig:Anorm_transient}(b) displays the norm of snapshots of the homogeneous system response over time, along with a reference line representing $e^{\lambda_l t}$. The natural decay rate aligns with the reference line, as expected, ensuring that the transient run will ultimately converge to the least-damped Floquet mode. Additionally, the system is not singular, as the transient dynamics never reach a naturally stable state, and none of the Floquet exponents have a zero real value.

Before proceeding to the computation of harmonic resolvent modes and gains using RSVD-\Dt and RSVD-LU, the transient removal strategy is demonstrated, and computing the actions of $\bm H$ and $\bm H^*$ are validated for a given random forcing containing $\Omega_{q} \in \{-10, -9, ..., 9, 10\}\omega_f$, \ie $\omega \in [-1, 1]$ with $\Delta \omega = 0.1$. The responses have been computed up to the $20^{th}$ harmonic, \ie $\omega \in [-2, 2]$, using both time stepping and directly in Fourier space as outlined in \eqref{eqn:Hres_eqn}. We used a $4^{th}$ order Runge--Kutta (RK4) integration scheme and a time step of $dt = 0.001$ and $T_t = 15000$ for this purpose. 

\begin{figure}[h]
\centering
\includegraphics[width=\textwidth,height=\textheight,keepaspectratio]{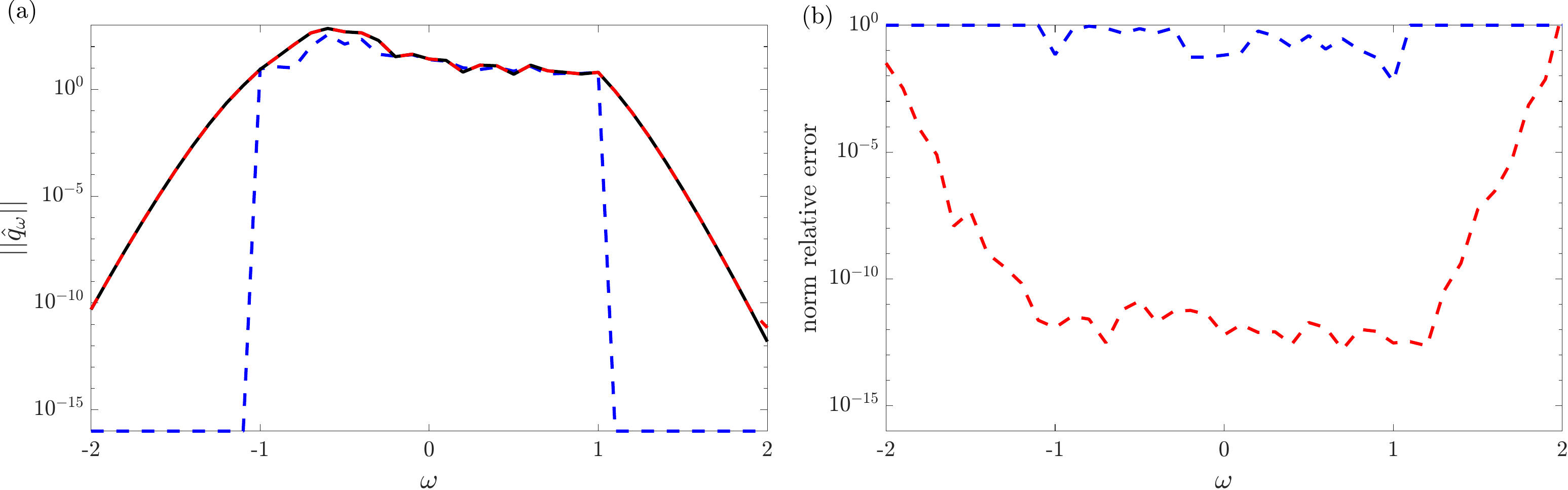}
\caption{Ginzburg-Landau test case: (a) the norms of the Fourier space response for $N_b = 3$ (black) and $N_b = 1$ (blue), along with the norm of the steady-state response obtained via time-stepping (red). (b) The relative errors of the norms, with the norms from Fourier space with $N_b = 3$ (black line in (a)) serving as the reference.}
\label{fig:validation_GL}
\end{figure}

Figure \ref{fig:validation_GL}(a) depicts the spectrum of response norms. In order to highlight the substantial difference in output generated by the harmonic resolvent operator around the base flow ($\Omega_{\bar q} = \{0\}\omega_f, N_b = 1$), the norm of the responses for the same input forcing is also presented. Figure \ref{fig:validation_GL}(b) displays the relative errors. The harmonic resolvent with $N_b = 1$ produces inaccurate results across the entire spectrum and has zero response for $|\omega| > 1$ since the modes at different frequencies are entirely decoupled. Therefore, neglecting $\hat{\bm A}_{p, \pm1}$ is impractical, as mentioned earlier, and induces significant alterations in the harmonic resolvent modes and gains. On the other hand, the time-stepping output exhibits almost machine-precision accuracy within $|\omega| \le 1$, and its accuracy diminishes as response norms decrease within $|\omega| > 1$. The validation of $\bm H^*$ follows a similar procedure but is omitted here for brevity.

The transient norm decay rate, as shown in figure \ref{fig:Anorm_transient}(b), follows $e^{-0.0021t}$. This indicates that approximately 40 periods are necessary for the transient norm to decay to 1\%. However, we can significantly reduce this duration by employing our transient removal strategy. To begin, we precompute the \edit{trial} basis by conducting a sufficiently long homogeneous simulation, and we find that a single mode suffices. Next, we integrate the orthogonalized basis for one period to obtain $\tilde{M}$ and complete the preprocessing step. 

\begin{figure}[h]
\centering
\includegraphics[width=\textwidth,height=\textheight,keepaspectratio]{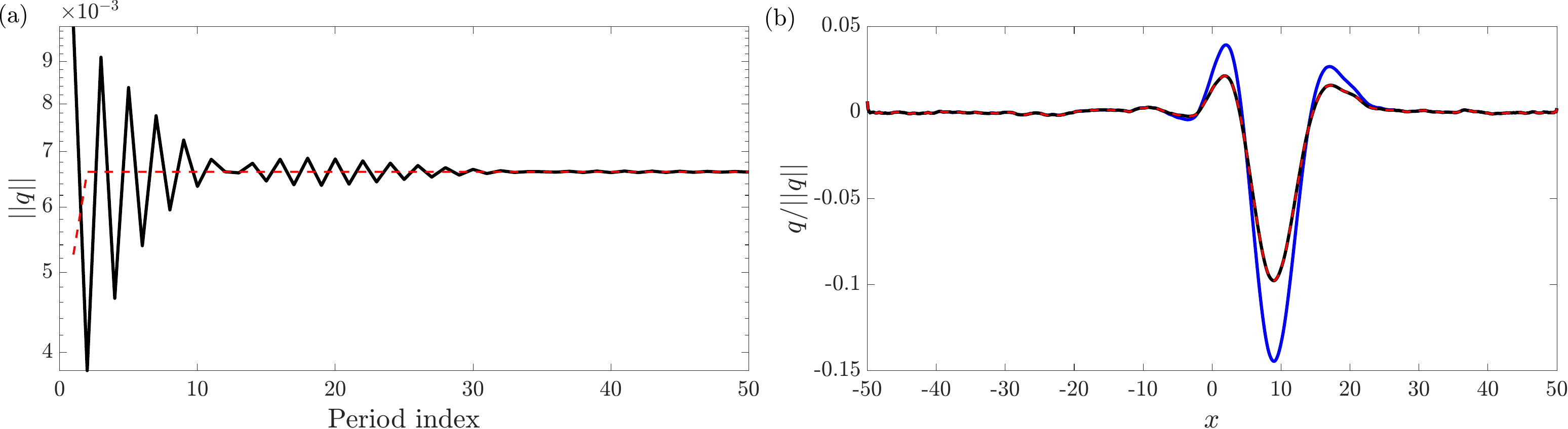}
\caption{Ginzburg-Landau test case: (a) the response norms before (black) and after (red) transient removal at the same phase, \ie snapshots with $\Delta t = T$ interval. (b) Comparison of the normalized response at the second period in blue, the $50^{th}$ period in black, and the second period after transient removal in red.}
\label{fig:TransientRemoval_GL}
\end{figure}

Employing the same forcing as the validation case, we obtain the responses in time at the identical phase (initiated at $t = 0$ or $\theta = 0$ with a time interval of $\Delta t = T$), and compare them with the solutions after the removal of transients. Figure \ref{fig:TransientRemoval_GL}(a) presents the norm of the responses, which illustrates the effectiveness of our strategy commencing from the second period. This plot suggests that after two periods using our transient removal strategy with Galerkin projection, we obtain an accurate steady-state snapshot which can be used as the synchronized initial condition for the forced equations. Figure \ref{fig:TransientRemoval_GL}(b) demonstrates that the snapshots ultimately align perfectly with the updated snapshot. Hence, using our strategy, a total of three periods are required -- two periods before transient removal and one period after the initial condition is obtained -- to acquire all $N_{\omega}$ steady-state snapshots, a 10-fold speed-up compared to the natural decay. A similar observation was made regarding the adjoint equations. In brief, our strategy can significantly reduce computational costs, potentially by one or more orders of magnitude, depending on the desired level of accuracy. 

\begin{figure}[h]
\centering
\includegraphics[width=\textwidth,height=\textheight,keepaspectratio]{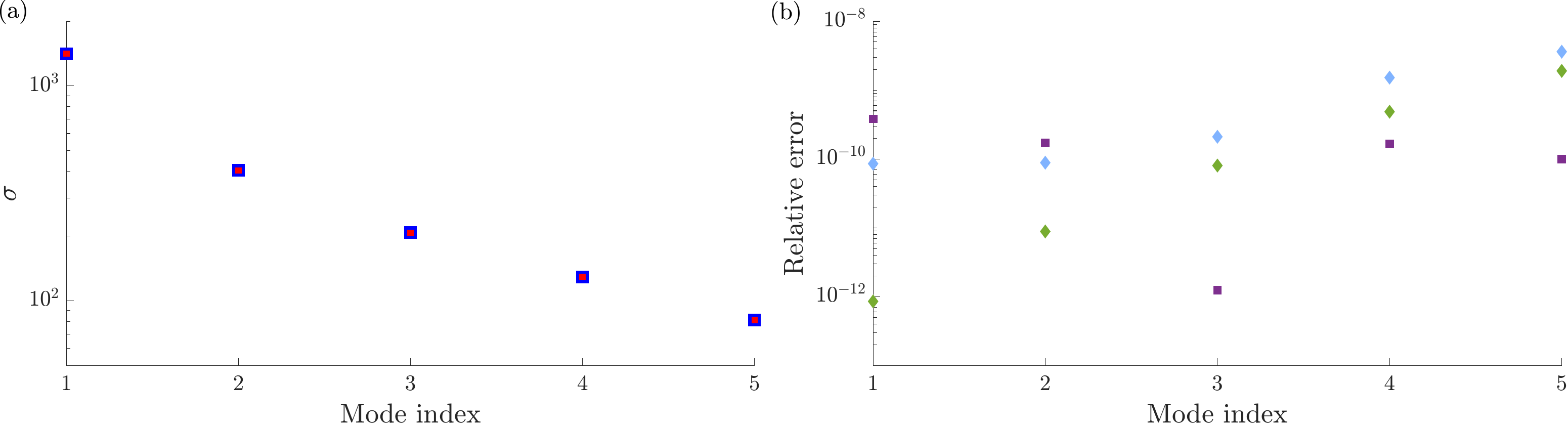}
\caption{Ginzburg-Landau test case: (a) the five leading gains using RSVD-\Dt (blue) and RSVD-LU (red). (b) The corresponding gain relative error (purple) and relative inner product errors \eqref{eqn:inner} between the response modes (cyan) and the forcing modes (green) computed via RSVD-\Dt and RSVD-LU.}
\label{fig:HModesGains_GL}
\end{figure}

Finally, RSVD-\Dt employing an RK4 integration scheme with $dt = 0.003$ and $T_t = 2T$ is carried out along with our efficient transient removal strategy to compute the harmonic resolvent modes for $k = 5$ and $q = 1$. To establish a reference, the harmonic resolvent modes are also computed using RSVD-LU with an equivalent number of test vectors and power iterations. Figure \ref{fig:HModesGains_GL}(a) illustrates that RSVD-\Dt computes the five leading gains identically to RSVD-LU. The gain relative error in figure \ref{fig:HModesGains_GL}(b) confirms that relative errors remain below $10^{-9}$ across all modes. The inner products of response and forcing modes, computed via RSVD-\Dt and RSVD-LU, demonstrate parallel directions as the relative errors remain below $10^{-8}$ across all modes. The inner-product error between two unit-norm vectors $\bm v_1$ and $\bm v_2$ is defined as 
\begin{equation}
e_{ip} = 1 - \langle \boldsymbol{\bm v_1}, \boldsymbol{v_2} \rangle.
\label{eqn:inner}
\end{equation}

\subsection{Flow over an airfoil}

\begin{figure}[h]
\centering
\includegraphics[width=\textwidth,height=\textheight,keepaspectratio]{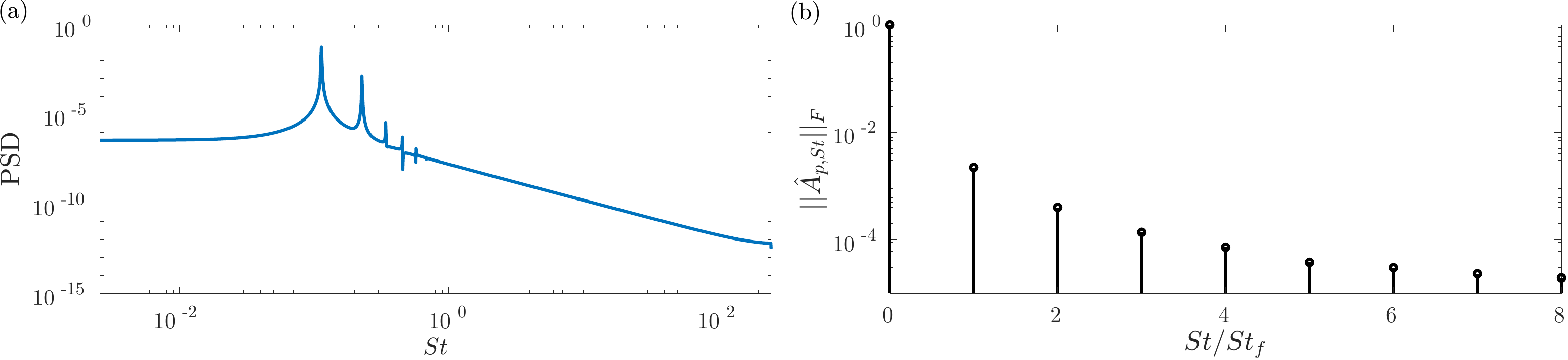}
\caption{Airfoil test case: (a) the PSD spectrum based on transverse velocity at $(x/L_c,y/L_c) = (3.0, -0.43)$. (b) The normalized Frobenius norm of $\hat{\bm A}_{p, St}$ up to the eighth harmonic.}
\label{fig:PSD_Fnorm}
\end{figure}

Our second test case consists of the flow over a NACA0012 airfoil at a low Reynolds number of $Re = 200$ and a steep angle of attack of $\alpha = 20^{\circ}$, which is dominated by periodic motions. This problem serves as a benchmark for computational cost and accuracy comparisons. A direct numerical simulation is conducted using the ``CharLES'' compressible flow solver. To mimic the incompressible simulations by \citet{Padovanetal20}, we set the Mach number to 0.05. The simulation employs a C-shaped mesh with a total of 62,000 cells. The leading edge of the airfoil is located at the origin $(x/L_c,y/L_c) = (0, 0)$, where $L_c = 1$ is the chord length. The computational domain spans $x/L_c \times y/L_c \in [-49,50] \times [-50,50]$. A constant time step of $\Delta t U_{\infty} /L_c = 6.88 \times 10^{-5}$ is used, corresponding to a  Courant-Friedrichs-Lewy number of 0.91, where $U_{\infty}$ is the inflow streamwise velocity. Integration continues for a sufficient duration until the flow reaches a periodic state. Figure \ref{fig:PSD_Fnorm}(a) displays the power spectral density computed from the transverse velocity at $(x/L_c,y/L_c) = (3.0, -0.43)$, where a pronounced vortex shedding occurs behind the trailing edge. The shedding frequency is $St_f = 0.114$, where the Strouhal number is defined as $St = \frac{\omega L_c sin(\alpha)}{2\pi U_{\infty}}$. The consistency between the dominant frequency identified in the power spectral density and obtained from our linearized code has been confirmed in our previous work \citep{Jungetal23}.

The linearized operators are constructed at 100 time points within $T = 2\pi / St_f \approx 55$. Performing a DFT on the $\bm A_i$, we generate 100 LNS operators $\hat{\bm A}_j$ in Fourier space to be used for harmonic resolvent analysis. The Frobenius norms $||\hat{\bm A}_{p, St}||_F$, depicted in figure \ref{fig:PSD_Fnorm}, indicate that $\bm A_p(t)$ can be accurately reconstructed using up to the 5$^{th}$ harmonic. Given that the linearized operator $\bm A_p(t)$ is real-valued for this problem, we retained only $\lfloor N_b/2 \rfloor + 1 = 6$ coefficients, \ie the zeroth and 5 positive harmonics. For time stepping, $dt = 0.0045$ is chosen to ensure the stability and accuracy of the RK4 integration scheme. The input and output perturbation frequency range spans up to the $7^{th}$ harmonic, \ie $N_{\omega} = 15$. The domain of interest is $x/L_c \times y/L_c \in [-4,12] \times [-2.5,2.5]$, identical to that in \citet{Padovanetal20}. We seek optimal forcing and response under the Chu energy norm \citep{Chuetal65}, which has been used in several previous studies \citep{Towneetal18, Schmidtetal18, HeidtColonius23}. The number of test vectors is set to $k = 5$, and $q = 2$ power iteration proves sufficient for the convergence of both gains and modes.

\begin{figure}[h]
\centering
\includegraphics[width=\textwidth,height=\textheight,keepaspectratio]{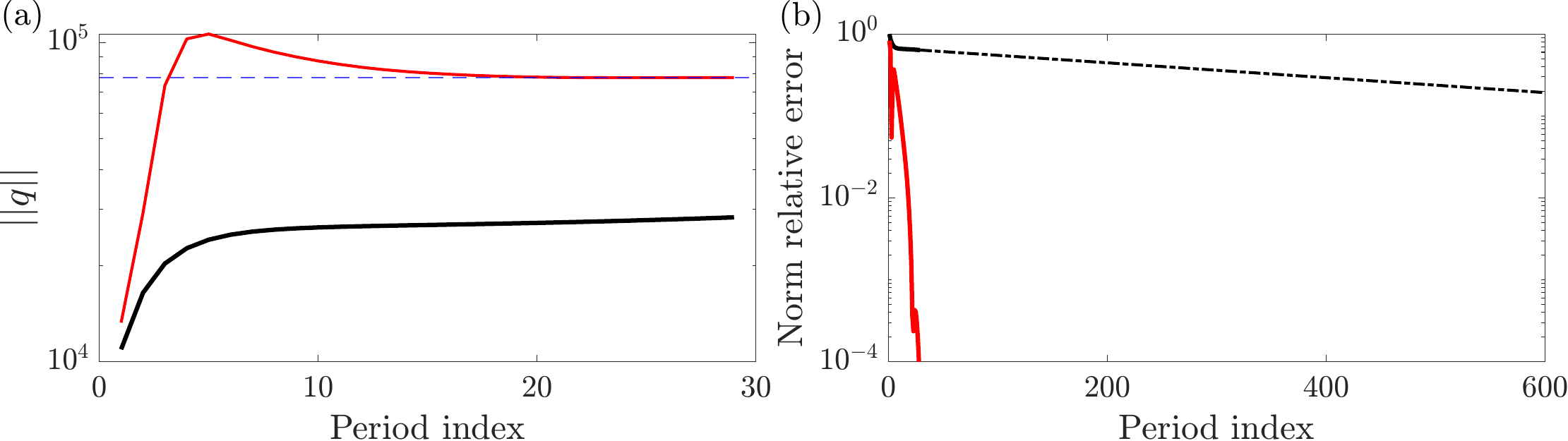}
\caption{Airfoil test case: (a) the response norms before (black) and after (red) transient removal, and the true steady-state norm (blue), at the same phase, \ie snapshots with $\Delta t = T$ interval, for a random forcing. (b) The relative error between the steady-state response norm and the response norms before (black) and after (red) transient removal.}
\label{fig:TransientRemoval_airfoil}
\end{figure}

A homogeneous simulation of the time-periodic system exhibits a slow decay rate, so we employ the transient removal strategy. Three snapshots constitute the \edit{trial} basis, obtained after a sufficiently long time interval where most initial transients are eliminated. A random forcing, encompassing the set $\Omega_{q} \in \{-7, -6, ..., 6, 7\}St_f$ frequencies, is applied to the LNS equations for over 30 periods. The snapshot norms slowly approach towards the steady-state norm but remain far from converging, as illustrated in figure \ref{fig:TransientRemoval_airfoil}. Utilizing our strategy, the relative norm error falls below the 1\% threshold after 20-30 periods in most cases with different random forcings. The efficacy of our strategy is further emphasized in figure \ref{fig:TransientRemoval_airfoil}(b), where the 1\% relative error is achieved before 10 periods for the optimal forcing (from RSVD-\Dt output). Ultimately, we set $T_t = 29T$ and conducted our simulations for a total duration of 30 periods to compute the actions of $\bm H$ and $\bm H^*$. Extrapolating the natural decay rate suggests that, in the absence of our transient-removal strategy,  the norms reach the 1\% threshold after around 2000 periods, demonstrating over a 60-fold acceleration in time stepping and overall algorithm performance with our strategy. This observation holds true for the adjoint equations as well.

\begin{figure}[h]
\centering
\includegraphics[width=\textwidth,height=\textheight,keepaspectratio]{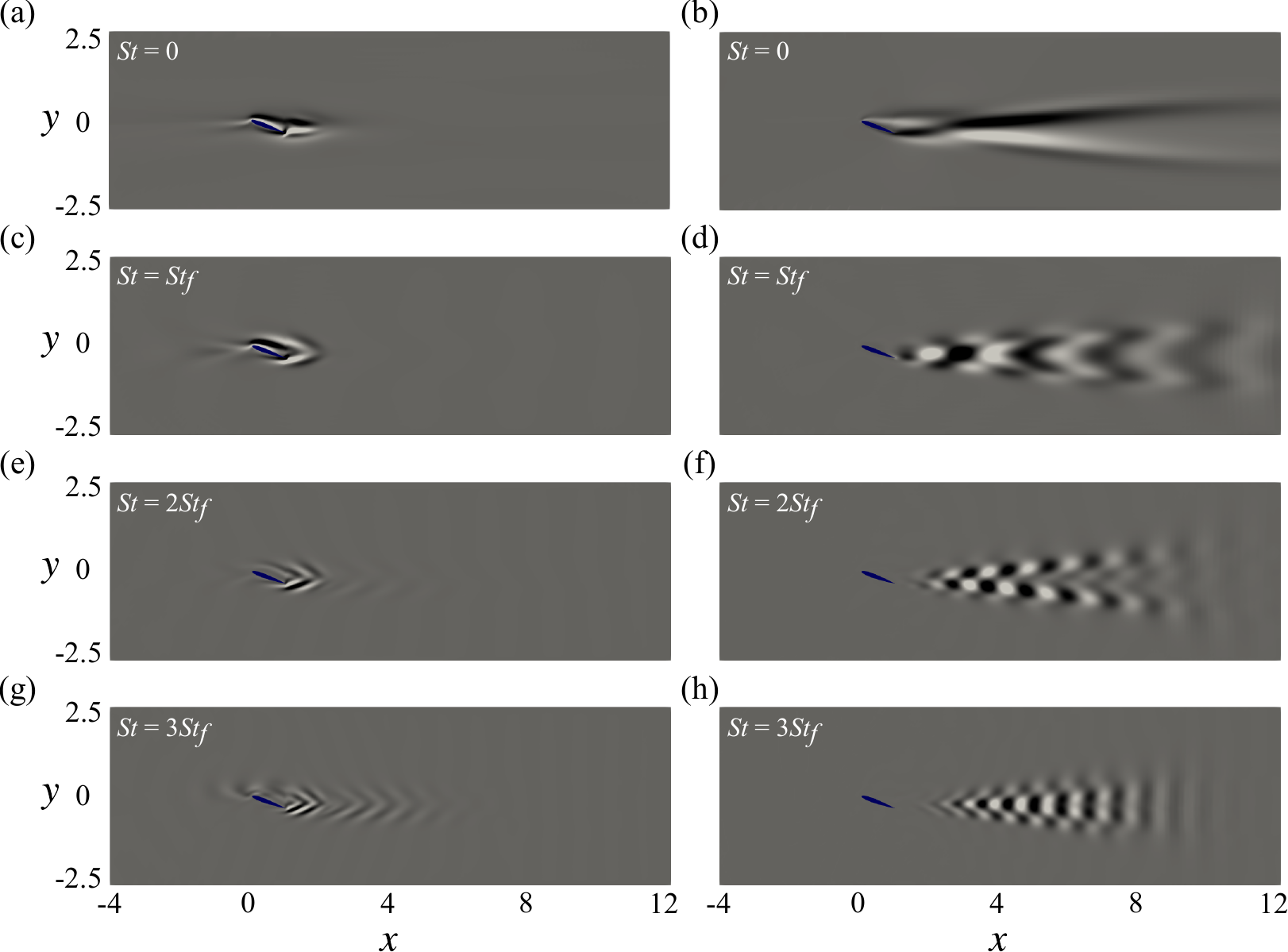}
\caption{Airfoil test case: real part of the vorticity field computed from (a, c, e, g) the input mode and (b, d, f, h) the output mode associated with the first suboptimal gain.}
\label{fig:HModes_airfoil}
\end{figure}

\edit{The optimal forcing and response modes (apart from the phase shift mode) are forcing and response modes are shown in figure \ref{fig:HModes_airfoil}. The vorticity patterns of the first output mode of the harmonic resolvent closely predict vortical structures observed in simulations that are induced by sinusoidal perturbations to the periodic base flow across all frequencies\citep{Padovanetal20}. This close match is attributed to the low-rank nature of the harmonic resolvent operator of the airfoil. This agreement also implies that regardless of the forcing type, resultant flow perturbations resemble the suboptimal output modes. Moreover, the proximity of forcing modes to the airfoil suggests sensitivity to perturbations near it. The significance of harmonic resolvent analysis lies in its ability to uncover flow structures that would remain hidden when linearizing around the temporal mean. It reveals the intricate interplay between different frequencies, the base frequency and its harmonics, which collectively contribute to the observed flow patterns during real perturbation analysis.}

\begin{figure}[h]
\centering
\includegraphics[width=\textwidth,height=\textheight,keepaspectratio]{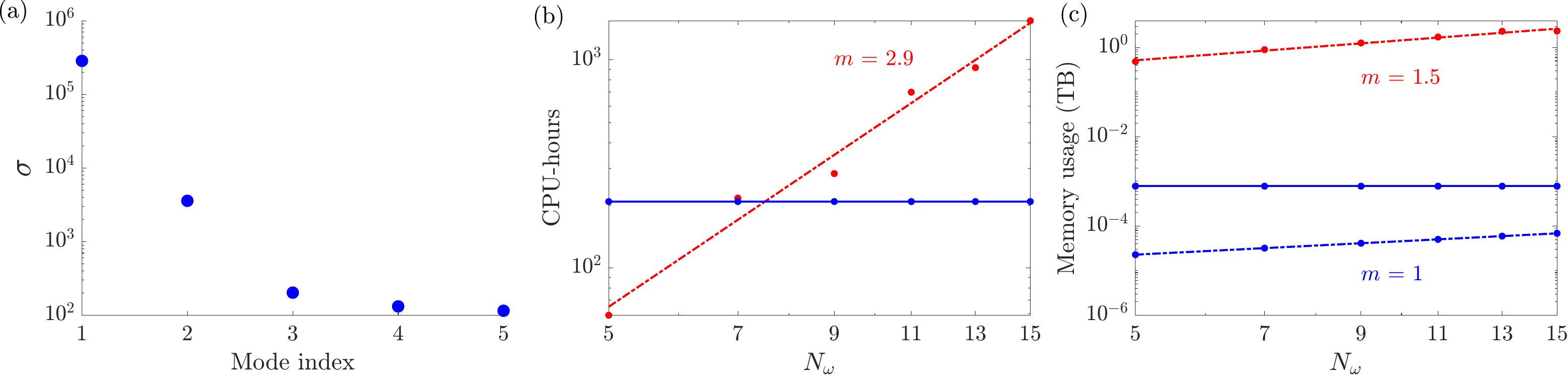}
\caption{Airfoil test case: (a) the five leading gains using RSVD-\Dt. (b) The CPU-hour, and (c) memory usage scaling of RSVD-LU (red) and RSVD-\Dt (blue) to compute the action of $\bm H$ onto a vector, \ie $k = 1$. The memory usage of RSVD-\Dt is decomposed into memory required to store LNS operators (solid) and forcing and response matrices (dashed).}
\label{fig:HGains_airfoil}
\end{figure}

The results from RSVD-\Dt closely align with the existing data obtained from RSVD-LU by \citet{Padovanetal20}, despite variations in the CFD solver, boundary conditions, domain setup, and energy norm. The gain plot in figure \ref{fig:HGains_airfoil}(a) exhibits a similar pattern to \citet{Padovanetal20}, with the exception that we have retained the optimal gain associated with the phase shift (the exact values differ due to differences in the problem setup). We noted that the leading mode and gain, \ie the mode associated with the phase shift, converged without power iterations due to a substantial two-orders-of-magnitude gap between the optimal and suboptimal gains.

From a computational standpoint, a fair comparison between the RSVD-LU and RSVD-$\Delta t$ algorithms is feasible when both utilize the same parameters. However, the use of $N_b = 11$ proves to be excessively memory-intensive for the RSVD-LU algorithm, exceeding the available 3.5 TB of memory on our cluster. The CPU and memory scaling plots in figure \ref{fig:HGains_airfoil}(b) and (c), respectively, depict the LU decomposition of $\bm T$ with $N_b = 3$, maintaining a constant $N$ while varying the number of blocks $N_{\omega}$ from 5 to 15 with $\Delta N_{\omega} = 2$. On the same plots, we present the cost of computing the action of $\bm H$ on a vector using time-stepping. The total duration is set to 30 periods to obtain an accurate solution. We only vary $N_{\omega}$ while keeping $N_b = 3$, the overall cost of RSVD-\Dt remains constant and unaffected by changes in $N_{\omega}$. This implies that the creation of LNS operators and the time-stepping process incur significantly higher costs compared to the transformations between Fourier space and the time domain for forcing and response. The memory of RSVD-$\Delta t$ to store LNS operators remains independent of $N_{\omega}$, as expected (see $\S$\ref{sec:mem}), while storing $\bm Q$ and $\bm F$ matrices grows linearly with the number of frequencies $N_{\omega}$.

RSVD-LU is estimated to require 1543 CPU-hours for the LU decomposition of $\bm T$, and an additional 15 CPU-hours for solving the LU-decomposed system for each test vector, yielding a total cost of $\text{CPU}_\text{{RSVD-LU}} \approx 1543 + 15 \times 5 \times 2 \times 2 = 1843$ CPU-hours for $k = 5, q = 1$. On the other hand, employing RSVD-\Dt with the aforementioned time-stepping parameters and $N_b = 11$ and $N_{\omega} = 15$ incurs a cost of approximately 7 CPU-hours per period for a test vector, resulting in a total cost of $\text{CPU}_\text{{RSVD-$\Delta t$}} \approx 30 \times 7 \times 5 \times 2 \times 2 = 4200$ CPU-hours. The cost of QR decomposition and the final SVD are ignored as they are orders of magnitude faster. In terms of memory consumption, RSVD-LU peaks at $\text{RAM}_{\text{RSVD-LU}} \approx 2036 GB$ for the LU decomposition of $\bm T$. RSVD-\Dt, on the other hand, stores $\lfloor N_b/2 \rfloor + 1 = 6$ sparse matrices, each of size 0.4 GB, and three dense matrices of size $Nk(\lfloor N_{\omega}/2 \rfloor + 1) = 313630 \times 5 \times 8$, totaling $\text{RAM}_\text{{RSVD-$\Delta t$}} \approx 0.4 \times 6 + 0.185 \times 3 \approx 3 GB$. This translates to a memory saving of approximately three orders of magnitude, even with an unbalanced number of base frequencies. Overcoming the memory consumption hurdle, which is typically the limiting factor in practice, RSVD-\Dt emerges as a viable tool for analyzing larger-scale flows compared to RSVD-LU.

\section{Conclusions} \label{sec:conclusion}

This paper presents an extension of the RSVD-\Dt algorithm that was originally developed for resolvent analysis of steady base flows to handle harmonic resolvent analysis of periodic flows. Specifically, we demonstrate that the time-stepping technique employed within RSVD-\Dt can effectively replace the actions of $\bm H$ and its complex conjugate $\bm H^*$ in Fourier space. In terms of CPU cost, the RSVD-\Dt algorithm exhibits a scaling of $O(N)$ for both resolvent and harmonic resolvent analyses. This offers a significant advantage, considering that the LU decomposition of $\bm T$ scales as $O((N_{\omega}N)^c)$ with $c \ge 1.5$. Regarding memory usage, our algorithm only stores relevant matrices in Fourier space and exhibits $O(NN_{\omega})$ scaling; in contrast, the memory peak of LU decomposition of $\bm T$ empirically scales with $O((N_{\omega}N)^{1.5})$ or worse. One difference between applying RSVD-\Dt to resolvent and harmonic resolvent analyses lies in the necessity to update the LNS operators during the time-stepping process for the latter. These operators are efficiently generated on the fly from their $N_b$ Fourier components.

The error sources in our algorithm align with those in resolvent analysis, extensively investigated in \citet{Farghadanetal23}. A unique contribution of this paper is the introduction of a novel transient removal strategy for harmonic resolvent analysis. While reminiscent of our approach for resolvent operators, the challenge in harmonic resolvent analysis lies in the intertwining of all retained frequencies. Our strategy takes advantage of the differing evolution of the steady-state and transient components of the response and Petrov-Galrkin or Galerkin projections. 

A validation of RSVD-\Dt against RSVD-LU is conducted using a periodic Ginzburg-Landau system. Additionally, we verify the governing role of the Floquet exponent in determining the decay rate of the least-damped mode of the system. Extending the application of our algorithm, we analyze a two-dimensional flow passing an airfoil, providing insights into the CPU and memory complexities. Despite dealing with a mid-sized flow scenario, a substantial memory gap persists between RSVD-LU and RSVD-\Dt algorithms. The computed forcing and response modes corresponding to the first suboptimal gain closely resemble those obtained by \citet{Padovanetal20}. The adaptation of the transient removal strategy for these test cases significantly enhanced the performance of RSVD-$\Delta t$, resulting in a 10-fold speed-up for the Ginzburg-Landau problem and a 60-fold speed-up for the airfoil to reach a 1\% relative error. The speed-up is even greater for lower error tolerances.

Moreover, our algorithm exhibits versatility, accommodating non-identity weight, input, and output matrices. This capability extends to computing modes for the modified harmonic resolvent operator, and in particular, the analysis of cross-frequency amplification mechanisms via $\bm H_{\omega_2, \omega_1}$. Our algorithm tackles the computation of subharmonic resolvent modes if desired. Implementation of the RSVD-\Dt algorithm within Petsc \citep{Balayetal19} and Slepc \citep{Hernandezetal05} environments leverage parallel computations for enhanced efficiency. Our time-stepping approach allows for a matrix-free application. This can be implemented using any code equipped with linear direct and adjoint capabilities, bypassing the explicit formation of $\bm T$ \citep{dePandoetal12, Martinietal21}. Finally, the time stepping within our algorithm can also be used along with other SVD algorithms, \eg Arnoldi and power iteration, if desired.  

In turbulent flows, where the state dimension $N$ can become very large due to higher resolution requirements, CPU cost and memory requirements can swiftly impose constraints on the applicability of RSVD-LU. This limitation applies even to steady problems and is further exacerbated for harmonic resolvent analysis due to the inflated frequency-domain operators necessitated by triadic frequency coupling. Our algorithm circumvents this issue by strategically operating in the time domain, avoiding the need for limiting operations like LU decomposition, leading to linear cost scaling. Because of this key distinction, the RSVD-\Dt could enable harmonic resolvent analysis of previous intractable turbulent flows.  

\section*{Acknowledgements}
This work was funded in part by Air Force Office of Scientific Research (AFOSR) grant no. FA9550-20-1-0214 and by a Catalyst Grant from the Michigan Institute for Computational
Discovery and Engineering (MICDE). We acknowledge the University of Michigan's Great Lakes cluster for providing the essential computational resources to conduct all the simulations in this study.

\begin{appendices}

\section{RSVD-\texorpdfstring{$\Delta t$}{deltat} for the subharmonic resolvent operator}\label{appA}

In $\S$\ref{sec:sub}, we provided an overview of subharmonic resolvent analysis. In this appendix, we briefly outline the application of RSVD-\Dt in computing subharmonic resolvent modes and gains. Consider the frequency of interest as $\gamma \in \Omega_{\gamma}$. This set specifies the perturbation frequency, while the base flow frequency content is confined to $\Omega_{\bar q}$.

To compute the actions of $\bm H$ and $\bm H^*$ using time-stepping, we must adhere to both the base flow frequency, enforcing a duration of $T = 2\pi/\omega_f$, and the perturbation frequency, enforcing a duration of $T_p = 2\pi/\gamma$, in order to obtain $N_{\omega}$ steady-state solutions to \eqref{eqn:linsys_harmonic}. Thus, we need to integrate for $T_{sub}$ such that $T_{sub}/T \in \mathbb{N}$ and $T_{sub}/T_p \in \mathbb{N}$, during which $N_{\omega}$ equidistant snapshots are saved. For instance, if we consider $\gamma = \omega_f/5$, requiring $T_p = 2\pi/(\omega_f/5) = 5T$, integrating for $T_{sub} = 5T$ is sufficient to obtain the steady-state solutions. All the other steps remain the same as introduced in Algorithm \ref{alg:alg_RSVDt}.

\end{appendices}

\bibliographystyle{unsrtnat}
\bibliography{hRSVDt}  

\begin{thebibliography}{41}
\providecommand{\natexlab}[1]{#1}
\providecommand{\url}[1]{\texttt{#1}}
\expandafter\ifx\csname urlstyle\endcsname\relax
  \providecommand{\doi}[1]{doi: #1}\else
  \providecommand{\doi}{doi: \begingroup \urlstyle{rm}\Url}\fi

\bibitem[Padovan et~al.(2020)Padovan, Otto, and Rowley]{Padovanetal20}
A.~Padovan, S.~E. Otto, and C.~W. Rowley.
\newblock Analysis of amplification mechanisms and cross-frequency interactions
  in nonlinear flows via the harmonic resolvent.
\newblock \emph{Journal of Fluid Mechanics}, 900, 2020.

\bibitem[Taira et~al.(2017)Taira, Brunton, Dawson, Rowley, Colonius, McKeon,
  Schmidt, Gordeyev, Theofilis, and Ukeiley]{Tairaetal17}
K.~Taira, S.~L. Brunton, S.~T.~M. Dawson, C.~W. Rowley, T.~Colonius, B.~J.
  McKeon, O.~T. Schmidt, S.~Gordeyev, V.~Theofilis, and L.~S. Ukeiley.
\newblock Modal analysis of fluid flows: An overview.
\newblock \emph{AIAA Journal}, 55\penalty0 (12):\penalty0 4013--4041, 2017.

\bibitem[Towne et~al.(2018)Towne, Schmidt, and Colonius]{Towneetal18}
A.~Towne, O.~T. Schmidt, and T.~Colonius.
\newblock Spectral proper orthogonal decomposition and its relationship to
  dynamic mode decomposition and resolvent analysis.
\newblock \emph{Journal of Fluid Mechanics}, 847:\penalty0 821--867, 2018.

\bibitem[Dawson and McKeon(2019)]{DawsonMcKeon19}
S.~T.~M. Dawson and B.~J. McKeon.
\newblock On the shape of resolvent modes in wall-bounded turbulence.
\newblock \emph{Journal of Fluid Mechanics}, 877:\penalty0 682--716, 2019.

\bibitem[Sipp and Marquet(2013)]{SippMarquet13}
D.~Sipp and O.~Marquet.
\newblock Characterization of noise amplifiers with global singular modes: the
  case of the leading-edge flat-plate boundary layer.
\newblock \emph{Theoretical and Computational Fluid Dynamics}, 27\penalty0
  (5):\penalty0 617--635, 2013.

\bibitem[Pickering et~al.(2021)Pickering, Towne, Jordan, and
  Colonius]{Pickeringetal21noise}
E.~Pickering, A.~Towne, P.~Jordan, and T.~Colonius.
\newblock Resolvent-based modeling of turbulent jet noise.
\newblock \emph{The Journal of the Acoustical Society of America}, 150\penalty0
  (4):\penalty0 2421--2433, 2021.

\bibitem[Troy and Koseff(2005)]{TroyKoseff05}
C.~D. Troy and J.~R. Koseff.
\newblock The instability and breaking of long internal waves.
\newblock \emph{Journal of Fluid Mechanics}, 543:\penalty0 107--136, 2005.

\bibitem[Giannenas et~al.(2022)Giannenas, Laizet, and Rigas]{Giannenasetal22}
A.~E. Giannenas, S.~Laizet, and G.~Rigas.
\newblock Harmonic forcing of a laminar bluff body wake with rear pitching
  flaps.
\newblock \emph{Journal of Fluid Mechanics}, 945:\penalty0 A5, 2022.

\bibitem[Farghadan and Arzani(2019)]{FarghadanArzani19}
A.~Farghadan and A.~Arzani.
\newblock The combined effect of wall shear stress topology and magnitude on
  cardiovascular mass transport.
\newblock \emph{International Journal of Heat and Mass Transfer}, 131:\penalty0
  252--260, 2019.

\bibitem[Wereley(1990)]{Wereley90}
N.~M. Wereley.
\newblock \emph{Analysis and control of linear periodically time varying
  systems}.
\newblock PhD thesis, Massachusetts Institute of Technology, 1990.

\bibitem[Lin et~al.(2023)Lin, Tsai, and Tsai]{Linetal23}
C.-T. Lin, M.-L. Tsai, and H.-C. Tsai.
\newblock Flow control of a plunging cylinder based on resolvent analysis.
\newblock \emph{Journal of Fluid Mechanics}, 967:\penalty0 A41, 2023.

\bibitem[Franceschini et~al.(2022)Franceschini, Sipp, Marquet, Moulin, and
  Dandois]{Franceschinietal22}
L.~Franceschini, D.~Sipp, O.~Marquet, J.~Moulin, and J.~Dandois.
\newblock Identification and reconstruction of high-frequency fluctuations
  evolving on a low-frequency periodic limit cycle: application to turbulent
  cylinder flow.
\newblock \emph{Journal of Fluid Mechanics}, 942:\penalty0 A28, 2022.

\bibitem[Heidt and Colonius(2023)]{HeidtColonius23}
L.~Heidt and T.~Colonius.
\newblock Spectral proper orthogonal decomposition of harmonically forced
  turbulent flows.
\newblock \emph{arXiv preprint arXiv:2305.05628}, 2023.

\bibitem[Wu et~al.(2022)Wu, Meneveau, Mittal, Padovan, Rowley, and
  Cattafesta]{Wuetal22}
W.~Wu, C.~Meneveau, R.~Mittal, A.~Padovan, C.~W. Rowley, and L.~Cattafesta.
\newblock Response of a turbulent separation bubble to zero-net-mass-flux jet
  perturbations.
\newblock \emph{Physical Review Fluids}, 7\penalty0 (8):\penalty0 084601, 2022.

\bibitem[Padovan and Rowley(2022)]{PadovanRowley22}
A.~Padovan and C.~W. Rowley.
\newblock Analysis of the dynamics of subharmonic flow structures via the
  harmonic resolvent: Application to vortex pairing in an axisymmetric jet.
\newblock \emph{Physical Review Fluids}, 7\penalty0 (7):\penalty0 073903, 2022.

\bibitem[Jin et~al.(2021)Jin, Symon, and Illingworth]{Jinetal21}
B.~Jin, S.~Symon, and S.~J. Illingworth.
\newblock Energy transfer mechanisms and resolvent analysis in the cylinder
  wake.
\newblock \emph{Physical Review Fluids}, 6\penalty0 (2):\penalty0 024702, 2021.

\bibitem[Herbert(1984)]{Herbert84}
T.~Herbert.
\newblock Analysis of the subharmonic route to transition in boundary layers.
\newblock In \emph{22nd Aerospace Sciences Meeting}, page~9, 1984.

\bibitem[Edgington-Mitchell et~al.(2021)Edgington-Mitchell, Wang, Nogueira,
  Schmidt, Jaunet, Duke, Jordan, and Towne]{Edgingtonetal21}
D.~Edgington-Mitchell, T.~Wang, P.~Nogueira, O.~Schmidt, V.~Jaunet, D.~Duke,
  P.~Jordan, and A.~Towne.
\newblock Waves in screeching jets.
\newblock \emph{Journal of Fluid Mechanics}, 913:\penalty0 A7, 2021.

\bibitem[Kumar and Prakash(2022)]{KumarPrakash22}
C.~Kumar and A.~Prakash.
\newblock Effect of mass injection on secondary instability of hypersonic
  boundary layer over a blunt cone.
\newblock \emph{Physics of Fluids}, 34\penalty0 (6), 2022.

\bibitem[Moarref et~al.(2013)Moarref, Sharma, Tropp, and McKeon]{Moarrefetal13}
R.~Moarref, A.~S. Sharma, J.~A. Tropp, and B.~J. McKeon.
\newblock Model-based scaling of the streamwise energy density in
  high-{R}eynolds-number turbulent channels.
\newblock \emph{Journal of Fluid Mechanics}, 734:\penalty0 275--316, 2013.

\bibitem[Ribeiro et~al.(2020)Ribeiro, Yeh, and Taira]{Ribeiroetal20}
J.~H.~M. Ribeiro, C.-A. Yeh, and K.~Taira.
\newblock Randomized resolvent analysis.
\newblock \emph{Physical Review Fluids}, 5\penalty0 (3):\penalty0 033902, 2020.

\bibitem[Farghadan et~al.(2023)Farghadan, Martini, and Towne]{Farghadanetal23}
A.~Farghadan, E.~Martini, and A.~Towne.
\newblock Scalable resolvent analysis for three-dimensional flows.
\newblock \emph{arXiv preprint arXiv:2309.04617}, 2023.

\bibitem[Monokrousos et~al.(2010)Monokrousos, {\AA}kervik, Brandt, and
  Henningson]{Monokrousosetal10}
A.~Monokrousos, E.~{\AA}kervik, L.~Brandt, and D.~S. Henningson.
\newblock Global three-dimensional optimal disturbances in the blasius
  boundary-layer flow using time-steppers.
\newblock \emph{Journal of Fluid Mechanics}, 650:\penalty0 181--214, 2010.

\bibitem[Martini et~al.(2021)Martini, Rodr{\'\i}guez, Towne, and
  Cavalieri]{Martinietal21}
E.~Martini, D.~Rodr{\'\i}guez, A.~Towne, and A.~V.~G. Cavalieri.
\newblock Efficient computation of global resolvent modes.
\newblock \emph{Journal of Fluid Mechanics}, 919, 2021.

\bibitem[McKeon and Sharma(2010)]{MckeonSharma10}
B.~J. McKeon and A.~S. Sharma.
\newblock A critical-layer framework for turbulent pipe flow.
\newblock \emph{Journal of Fluid Mechanics}, 658:\penalty0 336--382, 2010.

\bibitem[Leclercq and Sipp(2023)]{LeclercqSipp23}
C.~Leclercq and D.~Sipp.
\newblock Mean resolvent operator of a statistically steady flow.
\newblock \emph{Journal of Fluid Mechanics}, 968:\penalty0 A13, 2023.

\bibitem[Halko et~al.(2011)Halko, Martinsson, and Tropp]{Halkoetal11}
N.~Halko, P.~Martinsson, and J.~A. Tropp.
\newblock Finding structure with randomness: Probabilistic algorithms for
  constructing approximate matrix decompositions.
\newblock \emph{SIAM Review}, 53\penalty0 (2):\penalty0 217--288, 2011.

\bibitem[Nyquist(1928)]{Nyquist28}
H.~Nyquist.
\newblock Certain topics in telegraph transmission theory.
\newblock \emph{Transactions of the American Institute of Electrical
  Engineers}, 47\penalty0 (2):\penalty0 617--644, 1928.

\bibitem[Towne et~al.(2022)Towne, Rigas, Kamal, Pickering, and
  Colonius]{Towneetal22}
A.~Towne, G.~Rigas, O.~Kamal, E.~Pickering, and T.~Colonius.
\newblock Efficient global resolvent analysis via the one-way {N}avier-{S}tokes
  equations.
\newblock \emph{Journal of Fluid Mechanics}, 948:\penalty0 A9, 2022.

\bibitem[Rigas et~al.(2021)Rigas, Sipp, and Colonius]{Rigasetal21}
G.~Rigas, D.~Sipp, and T.~Colonius.
\newblock Nonlinear input/output analysis: application to boundary layer
  transition.
\newblock \emph{Journal of Fluid Mechanics}, 911:\penalty0 A15, 2021.

\bibitem[Chomaz et~al.(1988)Chomaz, Huerre, and Redekopp]{Chomazetal88}
J.~M. Chomaz, P.~Huerre, and L.~G. Redekopp.
\newblock Bifurcations to local and global modes in spatially developing flows.
\newblock \emph{Physical Review Letters}, 60\penalty0 (1):\penalty0 25, 1988.

\bibitem[Hunt and Crighton(1991)]{HuntCrighton91}
R.~E. Hunt and D.~G. Crighton.
\newblock Instability of flows in spatially developing media.
\newblock \emph{Proceedings of the Royal Society of London. Series A:
  Mathematical and Physical Sciences}, 435\penalty0 (1893):\penalty0 109--128,
  1991.

\bibitem[Bagheri et~al.(2009)Bagheri, Henningson, Hoepffner, and
  Schmid]{Bagherietal09}
S.~Bagheri, D.~S. Henningson, J.~Hoepffner, and P.~J. Schmid.
\newblock Input-output analysis and control design applied to a linear model of
  spatially developing flows.
\newblock \emph{Applied Mechanics Reviews}, 62\penalty0 (2), 2009.

\bibitem[Chen and Rowley(2011)]{ChenRowley11}
K.~K. Chen and C.~W. Rowley.
\newblock H2 optimal actuator and sensor placement in the linearised complex
  {G}inzburg--{L}andau system.
\newblock \emph{Journal of Fluid Mechanics}, 681:\penalty0 241--260, 2011.

\bibitem[Cavalieri et~al.(2019)Cavalieri, Jordan, and
  Lesshafft]{Cavalierietal19}
A.~V.~G. Cavalieri, P.~Jordan, and L.~Lesshafft.
\newblock Wave-packet models for jet dynamics and sound radiation.
\newblock \emph{Applied Mechanics Reviews}, 71\penalty0 (2), 2019.

\bibitem[Jung et~al.(2023)Jung, Bhagwat, and Towne]{Jungetal23}
J.~Jung, R.~Bhagwat, and A.~Towne.
\newblock Resolvent-based estimation of laminar flow around an airfoil.
\newblock \emph{AIAA Paper $\#$2023-0077}, 2023.

\bibitem[Chu(1965)]{Chuetal65}
B.-T. Chu.
\newblock On the energy transfer to small disturbances in fluid flow (part i).
\newblock \emph{Acta Mechanica}, 1\penalty0 (3):\penalty0 215--234, 1965.

\bibitem[Schmidt et~al.(2018)Schmidt, Towne, Rigas, Colonius, and
  Br{\`e}s]{Schmidtetal18}
O.~T. Schmidt, A.~Towne, G.~Rigas, T.~Colonius, and G.~A. Br{\`e}s.
\newblock Spectral analysis of jet turbulence.
\newblock \emph{Journal of Fluid Mechanics}, 855:\penalty0 953--982, 2018.

\bibitem[Balay et~al.(2019)Balay, Abhyankar, Adams, Brown, Brune, Buschelman,
  Dalcin, Dener, Eijkhout, Gropp, et~al.]{Balayetal19}
S.~Balay, S.~Abhyankar, M.~Adams, J.~Brown, P.~Brune, K.~Buschelman, L.~Dalcin,
  A.~Dener, V.~Eijkhout, W.~Gropp, et~al.
\newblock \emph{PETSc users manual}.
\newblock Argonne National Laboratory, 2019.

\bibitem[Hernandez et~al.(2005)Hernandez, Roman, and Vidal]{Hernandezetal05}
V.~Hernandez, J.~E. Roman, and V.~Vidal.
\newblock Slepc: A scalable and flexible toolkit for the solution of eigenvalue
  problems.
\newblock \emph{ACM Transactions on Mathematical Software (TOMS)}, 31\penalty0
  (3):\penalty0 351--362, 2005.

\bibitem[de~Pando et~al.(2012)de~Pando, Sipp, and Schmid]{dePandoetal12}
M.~F. de~Pando, D.~Sipp, and P.~J. Schmid.
\newblock Efficient evaluation of the direct and adjoint linearized dynamics
  from compressible flow solvers.
\newblock \emph{Journal of Computational Physics}, 231\penalty0 (23):\penalty0
  7739--7755, 2012.

\end{thebibliography}

\end{document}